\documentclass{bmcart}

\usepackage{amsthm,amsmath}
\RequirePackage{natbib}
\usepackage[utf8]{inputenc} 
\usepackage{graphicx}

\usepackage{soul}

\startlocaldefs
\endlocaldefs

\begin{document}

\begin{frontmatter}

\begin{fmbox}
\dochead{Research}


\title{Length scales and scale-free dynamics of dislocations in dense solid solutions}


\author[
   addressref={aff1},                   
   corref={aff1},                       
   email={pgabor@caesar.elte.hu}   
]{\inits{GP}\fnm{G\'abor} \snm{P\'eterffy}}
\author[
   addressref={aff1},
]{\inits{PDS}\fnm{P\'eter D.} \snm{Isp\'anovity}}
\author[
   addressref={aff3},
]{\inits{MEF}\fnm{Michael E.} \snm{Foster}}
\author[
   addressref={aff3},
]{\inits{XWZ}\fnm{Xiaowang W.} \snm{Zhou}}
\author[
   addressref={aff2},
   email={ryan.sills@rutgers.edu} 
]{\inits{RBS}\fnm{Ryan B.} \snm{Sills}}


\address[id=aff1]{
  \orgname{Department of Materials Physics, E{\"o}tv{\"o}s Lor{\'a}nd University}, 
  \street{P\'azm\'any P. stny. 1/A}
  %
  \postcode{H-1117}
  \city{Budapest},                              
  \cny{Hungary}                                    
}

\address[id=aff2]{%
  \orgname{Department of Materials Science and Engineering, Rutgers University},
  \city{Piscataway},
  \state{New Jersey},
  \postcode{08854},
  \cny{USA}
}

\address[id=aff3]{%
  \orgname{Sandia National Laboratories},
  \city{Livermore},
  \state{California},
  \postcode{94550},
  \cny{USA}
}


\begin{artnotes}
\end{artnotes}

\end{fmbox}


\begin{abstractbox}

\begin{abstract} 
The fundamental interactions between an edge dislocation and a random solid solution are studied by analyzing dislocation line roughness profiles obtained from molecular dynamics simulations of $\rm Fe_{0.70}Ni_{0.11}Cr_{0.19}$ over a range of stresses and temperatures.
These roughness profiles reveal the hallmark features of a depinning transition.
Namely, below a temperature-dependent critical stress, the dislocation line exhibits roughness in two different length scale regimes which are divided by a so-called correlation length.
This correlation length increases with applied stress and at the critical stress (depinning transition or yield stress) formally goes to infinity.
Above the critical stress, the line roughness profile converges to that of a random noise field.
Motivated by these results, a physical model is developed based on the notion of coherent line bowing over all length scales below the correlation length.
Above the correlation length, the solute field prohibits such coherent line bow outs.
Using this model, we identify potential gaps in existing theories of solid solution strengthening and show that recent observations of length-dependent dislocation mobilities can be rationalized.
\end{abstract}




\end{abstractbox}
%

\end{frontmatter}



\section*{Introduction}

Solid solution strengthening is one of the most fundamental strengthening mechanisms for crystalline materials. 
Dating back 5000 years to the bronze age, the premise is quite simple: mix small amounts of a secondary metal into a primary metal to form a solid solution, known as an alloy, which is stronger than either pure metal.
It wasn't until the discovery of dislocations in the 1940s that the mechanism underlying solid solution strengthening was understood: solute atoms impede the motion of dislocation lines.
While qualitatively simple, theoretical descriptions of solid solution strengthening that are free of phenomenological parameters have been elusive.
With the advent of modern atomistic computing techniques and advanced scientific computing platforms, it is now possible to study solid solution strengthening at high fidelity and understand its fundamental nature. 

Classical theories for solid solution strengthening were developed by Fleischer, Friedel, Labusch, and Mott in the 1960s~\cite{mottMechanicalStrengthCreep1952,fleischerSubstitutionalSolutionHardening1963,freidelDislocations1964,labuschStatisticalTheorySolid1970,ardellPrecipitationHardening,argonStrengtheningMechanismsCrystal2008,varvenneSoluteStrengtheningRandom2017}.
These theories are now commonly referred to as ``strong pinning'' (Fleischer and Friedel) and ``weak pinning'' (Mott and Labusch), a distinction which refers to the nature by which the dislocation ``samples the solute field'' as it glides forward~\cite{ardellPrecipitationHardening,varvenneSoluteStrengtheningRandom2017}.
In strong pinning, the dislocation encounters each solute individually, overcoming a single solute obstacle at a time.
With weak pinning, the dislocation line instead encounters a diffuse solute field, and each point along the dislocation line is at all times responding to many solute interactions.
In the classical theories, this diffuse interaction is characterized by an ad hoc interaction length scale.
It has been shown that at most solute concentrations of interest for alloys (i.e., $> 0.1\,\,\rm at.\%$), weak pinning is the operative strengthening mode~\cite{leysonFriedelVsLabusch2013}.

Building on this classical work, Leyson et al.~\cite{leysonQuantitativePredictionSolute2010} recently developed a novel weak-pinning theory of solid solution strengthening.
In their theory, a dislocation line embedded in a field of solutes is assumed to form a quasi-sinusoidal geometry described by segment length $\zeta$ and amplitude $w$.
Values of the length scales $(\zeta,w)$ are then chosen such that the total energy of the dislocation is minimized, taking account of the elastic energy of the line and the interaction energies between each segment of the line and each solute atom. 
Finally, a temperature-dependent yield strength is derived using the obtained line geometry in conjunction with thermal activation theory.
The Leyson et al. theory provides a computationally tractable formalism for predicting solute strengthening while incorporating information from first-principles computations (density functional theory) of dislocation-solute interactions~\cite{leysonQuantitativePredictionSolute2010,leysonSoluteStrengtheningFirst2012}.
This theory has been used to successfully predict temperature-dependent yield strengths of Al, Zn, and high entropy alloys~\cite{varvenneTheoryStrengtheningFcc2016,varvenneSoluteStrengtheningRandom2017}.
The length scales $(\zeta,w)$ characterizing the dislocation line are the essential ingredients which enable the development of the theory.

Going beyond the yield strength (the stress below which dislocations are immobile), dislocation-solute interactions also critically influence the mobility of dislocation lines. 
Several authors have used molecular dynamics (MD) simulations to show the influence of solutes on the mobility curves (velocity as a function of shear stress) in face-centered cubic (FCC)~\cite{rodaryDislocationGlideModel2004,olmstedAtomisticSimulationsDislocation2005,marianMovingDislocationsDisordered2006,zhaoAtomicscaleDynamicsEdge2017,osetskyTwoModesScrew2019,sillsLineLengthDependentDislocationMobilities2020} and body-centered cubic (BCC)~\cite{itakuraEffectHydrogenAtoms2013,katzarovHydrogenEmbrittlementAnalysis2017} metals.
These studies show that as the solute concentration increases, the dislocation mobility is reduced (this is not always the case for screw dislocations in BCC metals, however~\cite{katzarovHydrogenEmbrittlementAnalysis2017}).
Several studies have shown that under some conditions, rather than moving smoothly in time the dislocation line occasionally arrests and is completely pinned in place~\cite{rodaryDislocationGlideModel2004,osetskyTwoModesScrew2019,sillsLineLengthDependentDislocationMobilities2020}.
The dislocation may then unpin and begin moving again.
Recently, this behavior was studied in detail and interpreted in terms of a length-dependent mobility; the authors showed that by increasing the line length $L$ a length-independent mobility could be recovered~\cite{sillsLineLengthDependentDislocationMobilities2020}.
It was argued on the basis of kinetic Monte Carlo simulations that this length-dependence is a direct result of the length scales which characterize the dislocation line's interactions with the solute field. 
However, a fundamental explanation derived from the details of these interactions is still lacking.

These works demonstrate the critical importance of the dislocation line geometry and its associated length scales in solid solution strengthening.
Recently, Zhai and Zaiser~\cite{zhai2019properties} studied the geometry of a dislocation line interacting with a spatially random force field (a model for a solid solution system) using discrete dislocation dynamics, demonstrating the hallmark behaviors of a so-called depinning transition~\cite{fisher1997statistics}.
They showed that the dislocation-force field interactions induced a stress-dependent characteristic length above which the roughness in the line geometry became uncorrelated; here, ``correlations'' refer to regions of the dislocation line which bow out in a coherent manner (i.e., are pinned).
They further showed that the characteristic length at zero-stress (deemed the ``pinning length'') scaled with the flow stress of the system.
In this work, we perform a similar study using line geometries obtained from molecular dynamics (MD) simulations of dislocation glide in a dense solid solution.
Our goal is to assess the geometry of dislocation lines over a range of stresses and temperatures in order to identify the scale parameters which govern their shape.
We show below that the trends in these parameters are intimately tied to the depinning transition, and help to rationalize the length-dependence in the dislocation mobility.
Our results also bring into question the concept of a unique set of length scales which govern solid solution strengthening.

The remainder of the manuscript is organized as follows.
First, because of its central role in our analysis, we give a brief review of the depinning transition of elastic manifolds, a concept which has been pervasively studied across many areas of physics.
In the Methods section, we discuss our simulation and analysis techniques.
In the Results section, we present the Hurst exponents $H$ and correlation length scales $\xi$ which we have extracted from the line geometries obtained from MD. 
Finally, in the Discussion section we consider the implications of our findings for theories of strengthening and dislocation mobility, and discuss future areas of study.

\section*{Depinning of elastic manifolds}

\begin{figure}[!ht]
  \caption{\csentence{The depinning transition.}
      Average velocity $v$ of the elastic manifold driven by the external force $F$. At the depinning point $F_\text{c}$ and $T=0$, $v$ starts to increase according to a power-law relationship.
      At finite temperature $T>0$, so-called thermal rounding leads to non-zero velocities $v>0$ for $F<F_\text{c}$.
      \label{fig:depinning_sketch}}
 \includegraphics[width=0.4\linewidth]{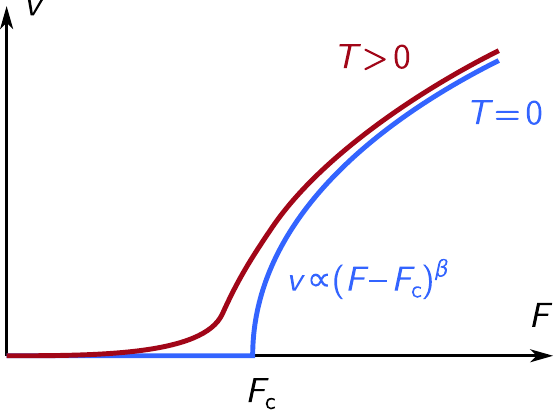}
\end{figure}

In this section a brief overview of the depinning transition \cite{fisher1985sliding, fisher1998collective} is given which, as will be shown below, properly describes the evolution of a dislocation line in the random alloy considered. The basic concept of this statistical physics approach is that an elastic manifold (typically a line or a surface) interacts with a static pinning field that exerts local pinning force on the manifold. External force $F$ is applied to the manifold which below a certain threshold force $F_\text{c}$ cannot move at zero temperature due to the pinning field. Whereas for $F>F_\text{c}$ the manifold depins, that is, it overcomes the restoring force of the pinning points and starts to move with a constant velocity $v$ (on average). The schematic plot of this phenomenon is shown in Fig.~\ref{fig:depinning_sketch}. At $T=0$, the $v(F)$ curve close to the depinning point $F_\text{c}$ can be usually approximated as
\begin{align}
    v(F) \propto (F-F_\text{c})^\beta, \qquad (F>F_\text{c})
    \label{eq:v_depinning}
\end{align}
where the velocity exponent $\beta>0$ is introduced. Examples of systems exhibiting depinning cover a wide range of physical phenomena: domain wall migration in ferromagnets \cite{zapperi1998dynamics}, motion of tectonic plates \cite{fisher1997statistics}, propagation of liquid fronts in porous materials \cite{martys1991critical}, and the propagation of crack fronts in heterogeneous materials \cite{daguier1997pinning}, just to mention a few.

The depinning takes place as a result of the competition between the restoring force due to the elastic properties (e.g., line tension) of the manifold which act to ``flatten'' it and the forces from the applied stress and pinning field which induce roughnening. This process is, perhaps counter-intuitively, highly non-local: as the force $F$ tends to $F_\text{c}$, numerous small rearrangements (or avalanches) of the manifold take place and, as a result, long range correlations build up in the shape of the manifold. Thus, the depinning transition is, in fact, a collective phenomenon and can be considered as a continuous phase transition.

\begin{figure}[!ht]
    \caption{\csentence{Roughness of a depinning line.} (a) Rough dislocation line with shape $y(x)$. The direction of the motion is parallel to the $y$ axis. Note that the pinning points are distributed randomly all over the $xy$ plane, in the figure only those are shown that are interacting with the dislocation line. (b) Roughness $R(l)$ at the depinning transition ($F=F_\text{c}$) with Hurst exponent $H$. Dashed lines denote the microscopic (e.g., atomic) and macroscopic (e.g., system size) limits, where the scaling breaks down.  (c) Roughness far from the depinning transition ($F \ne F_\text{c}$) with two Hurst exponents $H_1$ and $H_2$. (d) Expected behavior of the correlation length $\xi$ in the vicinity of the depinning transition.
    \label{fig:roughness}}
    \begin{picture}(0,0)
    \put(-15,70){\sffamily{(a)}}
    \end{picture}
    \includegraphics[width=0.45\linewidth]{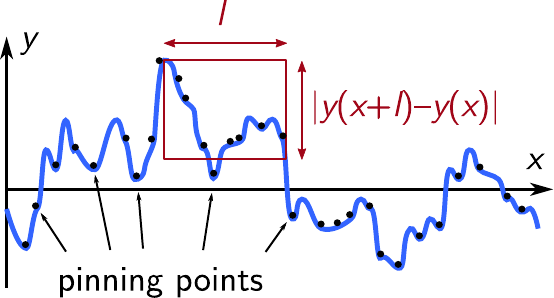} \vspace{0.5cm}\\
    \begin{picture}(0,0)
    \put(-15,90){\sffamily{(b)}}
    \end{picture}
    \includegraphics[width=0.3\linewidth]{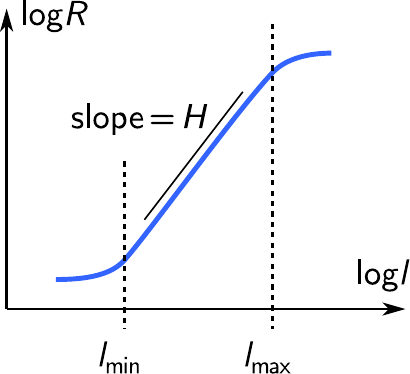}
    \hspace{0.5cm}
    \begin{picture}(0,0)
    \put(-15,90){\sffamily{(c)}}
    \end{picture}
    \includegraphics[width=0.3\linewidth, trim=0 0.0cm 0 0]{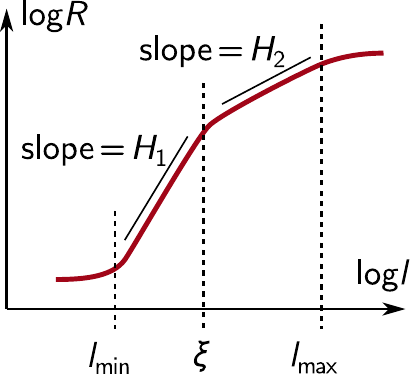}\\
    \vspace{0.6cm}
    \begin{picture}(0,0)
    \put(-15,90){\sffamily{(d)}}
    \end{picture}
    \includegraphics[width=0.3\linewidth, trim=0 0.0cm 0 0]{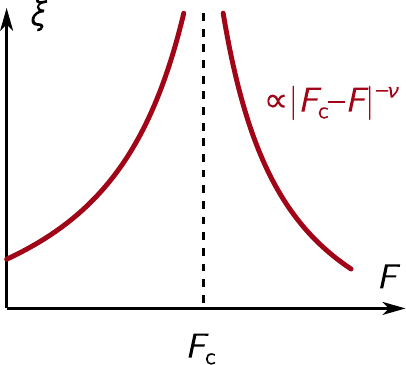}
\end{figure}

\begin{figure}[!ht]
  \caption{\csentence{Example line geometries with different Hurst exponents.} The curves represent examples of fractional Brownian motion, where correlations or anti-correlations are introduced between subsequent increments \cite{kroese2015spatial}. The Hurst exponent $H$ can be set in the model and $H=0.5$ corresponds to classical Brownian motion. (a) $H=0$, (b) $H=0.5$, (c) $H=1$. Each circular subplot magnifies the content of a small grid cell.
      \label{fig:hurst_exponent}}
 \includegraphics[width=0.9\linewidth]{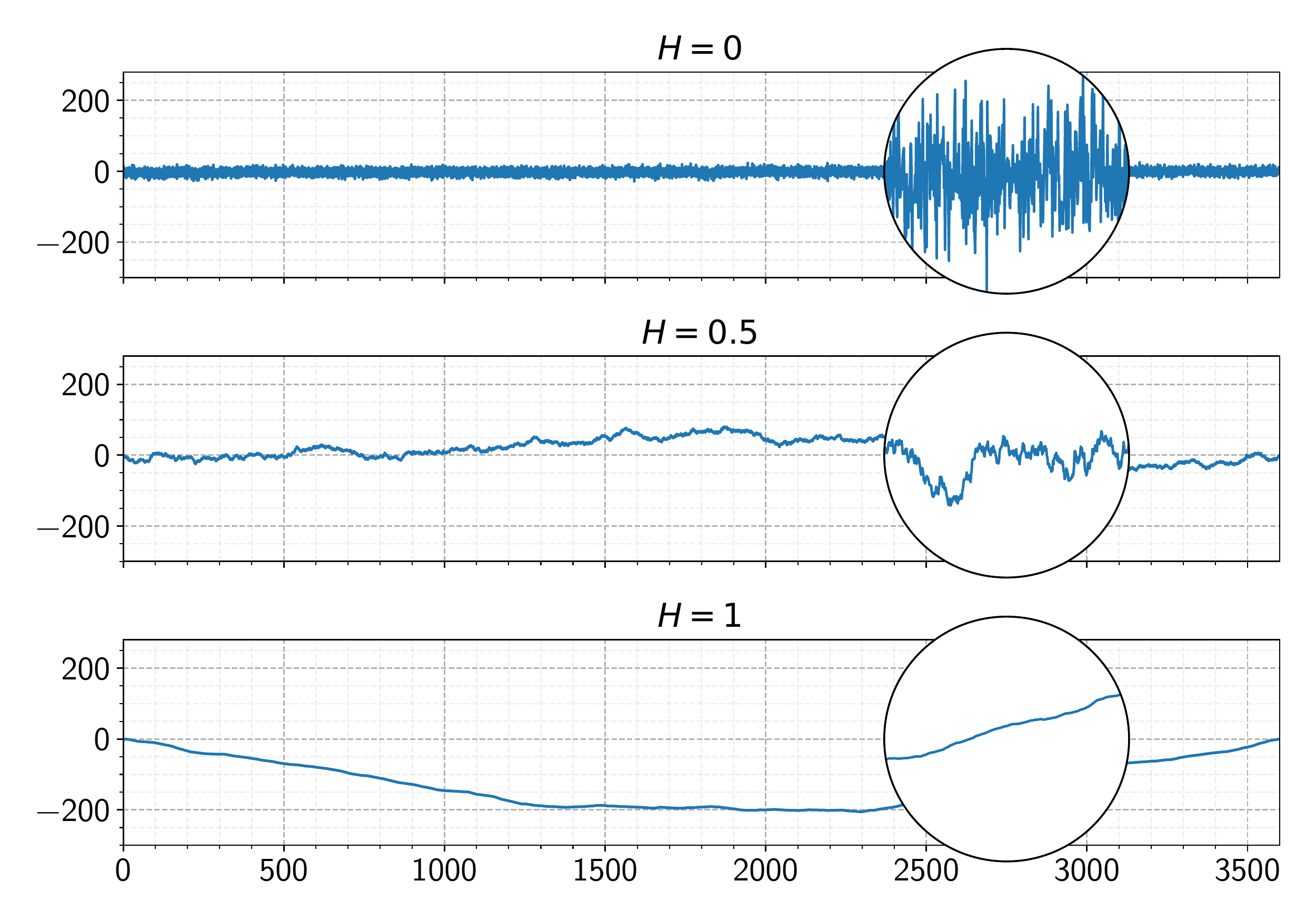}
 \begin{picture}(0,0)
     \put(-330,215){\sffamily{(a)}}
     \put(-330,145){\sffamily{(b)}}
     \put(-330,75){\sffamily{(c)}}
 \end{picture}
\end{figure}

The process of approaching the depinning point can be characterized by the so-called roughness function, which for line-like manifolds (e.g., dislocations) reads as
\begin{align}
    R(l) = \langle |y(x+l) - y(x)| \rangle, \label{eq:roughness}
\end{align}
where $y(x)$ represents the shape of the manifold (see sketch of Fig.~\ref{fig:roughness}a), and $\langle \cdot \rangle$ stands for averaging with respect to $x$, that is, along the direction of the manifold. (The force $F$ drives the manifold in the $+y$ direction.) This roughness function characterizes the average height difference of the curve at a typical distance $l$.  In the case of a gliding dislocation line, regions of the dislocation line bow out between dominant pinning points. As the applied stress $\sigma$ increases (giving a Peach-Koehler force $F=\sigma b$ where $b$ is the magnitude of the Burgers vector) and tends to the depinning (or yield) stress, the dislocation line overcomes weak pinning points and the bow out distance increases as the distance between the dominant pinning points increases. As a result, during depinning (i.e., when $F<F_c$) the roughness develops a power-law dependence as
\begin{align}
    R(l) \propto l^H,
    \label{eq:hurst}
\end{align}
with $H\geq0$ being the \emph{Hurst exponent} (Fig.~\ref{fig:roughness}b)\footnote{This exponent is often called the roughness exponent in the depinning literature.}. This means that the curve becomes self-affine with a fractal dimension $D=2-H$. Figure \ref{fig:hurst_exponent} presents examples of rough curves with Hurst exponents $H=0$, 0.5, and 1. The exponent of $H=0.5$ is equivalent to a one-dimensional Brownian motion where the increments of $y(x)$ between subsequent $x$ values are independent random variables. In the case of $H>0.5$ these increments are correlated (a positive step is likely followed by another positive step) whereas $H<0.5$ means anticorrelation between the steps (a positive step is likely followed by a negative step). In particular, $H=0$ corresponds to random (white) noise\footnote{The distinction between $H=0$ and $H=0.5$ is subtle.
The roughness in both cases can be thought of as random. 
For example, imagine that we construct a line connecting a set of points on a grid $(x_i,y_i)$ which are equally spaced in $x$. 
The position $y_i$ for each point is chosen at random.
However, the two cases differ in how the line is constructed.
In the case of $H=0$, each $y_i$ is chosen randomly in the interval $[y_{\rm min},y_{\rm max}]$.
The result is that the roughness of the line is anti-correlated, because a point on the line with $y_i$ at (or near) $y_{\rm max}$, for example, is most likely to be followed by a point $y_{i+1}<y_{\rm max}$.
Hence, an increase in $y$ position is likely followed by a decrease in $y$ position, the definition of anti-correlation.
In the case of $H=0.5$, the value for $y_{i+1}$ follows a \emph{random walk} process with $y_{i+1}=y_{i}+\Delta y$, where $\Delta y$ is chosen randomly in the range $[\Delta y_{\rm min},\Delta y_{\rm max}]$.
In this case the \emph{increment in $y$ position} is chosen randomly, rather than the $y$ position itself.
This leads to uncorrelated roughness.
}.

As the external force $F$ increases and tends to the depinning threshold $F_\text{c}$, the form of Eq.~(\ref{eq:hurst}) is gradually approached. In particular, this scaling behavior is only obeyed when $l$ is below a correlation length $\xi$. In the classical picture of depinning the correlation length is expected to diverge at the critical force $F_\text{c}$ as seen in Fig.~\ref{fig:roughness}(d) \cite{fisher1997statistics}. An exponent $\nu>0$ characterizing this divergence is introduced as
\begin{equation}
    \xi \propto |F - F_\text{c}|^{-\nu}.
    \label{eq:xi}
\end{equation}
This means that truly self-affine behavior (or, in other words, scale-free roughness) is only observed at the depinning point. At smaller and larger forces the length-scale $\xi$ appears, which usually separates two distinct regimes on the roughness plot each characterized by an individual Hurst exponent. Below the correlation length $\xi$, the curve is self-affine with a roughness exponent $H_1$ characteristic to the depinning transition. Above this scale thermal fluctuations and the local interaction between the pinning points and the manifold may introduce roughness with a different exponent $H_2$. It is noted, that the power-law dependence of the roughness $R$ cannot extend to all scales as there are physical barriers that pose a strict limit, such as the lattice constant and the size of the medium that the elastic manifold is embedded in, denoted by $l_{\rm min}$ and $l_{\rm max}$, respectively, in Fig.~\ref{fig:roughness}.

At $T>0$ the so-called thermal rounding is observed; that is, the $v(F)$ curve gets smooth close to the pinning point as shown in Fig.~\ref{fig:depinning_sketch}. Below the depinning point, thermal creep is observed, i.e., $v>0$, which has been extensively studied both theoretically \cite{chauve2000creep, purrello2017creep} and experimentally \cite{bustingorry2012thermal} for domain wall dynamics in ferromagnets.

In the context of dislocations, roughening due to pinning obstacles has been observed and studied quantitatively using discrete dislocation dynamics (DDD) simulations. In these models the motion of individual dislocation lines is tracked and the embedding medium is assumed to be an isotropically elastic continuum \cite{bulatov2006computer}. Zapperi and Zaiser constructed a lattice model to consider the gliding of a dislocation in a random field of forest dislocations \cite{zapperi2001depinning}. To model the effect of irradiation Hiratani and Zbib introduced stacking fault tetrahedra in copper that acted as barriers for dislocation motion \cite{hiratani2003dislocation}. Bakó et al.~investigated the hardening effect of ceramic dispersoids in steel by introducing impenetrable spherical obstacles in the slip plane of a single gliding dislocation \cite{bako2008dislocation}. 
Patinet et al.~\cite{patinetAtomicscaleAvalancheDislocation2011} analyzed edge dislocation depinning with molecular dynamics simulations of a random $\rm Ni_{0.9}Al_{\rm 0.1}$ alloy. They analyzed both the line roughness and the avalanche statistics associated with glide motion.
Finally, Zhai and Zaiser introduced a random pinning field that acted on the dislocation line to account for the  random distortion field of high entropy alloys \cite{zhai2019properties}. In all of these simulations scale-free roughness was found to develop as the yield stress was approached with Hurst (or roughness) exponents of $1.0$ \cite{zapperi2001depinning}, $0.8-0.85$\cite{hiratani2003dislocation}, $0.96$ \cite{bako2008dislocation}, $0.85$~\cite{patinetAtomicscaleAvalancheDislocation2011}, and $1.0$ \cite{zhai2019properties}, respectively. The effect of thermal noise in the absence of solutes was also studied by Zhai and Zaiser by adding Langevin forces that act on the dislocation lines~\cite{zhai2019properties}. It was found both theoretically and numerically that such thermal noise leads to a Hurst exponent of $H=0.5$. This was also found in MD simulations of dislocation fluctuations in pure Al (when $l >10\,\rm\AA$)~\cite{geslin2018thermal}.

The concept of depinning has also been introduced more generally for dislocation ensembles \cite{zaiser2006scale, friedman2012statistics, tsekenis2013determination}. It was found, however, with 2D and 3D DDD simulations that dislocation ensembles without a static pinning field do not undergo depinning, but instead belong to a different class of systems exhibiting scale-free fluctuations \cite{ispanovity2014avalanches, lehtinen2016glassy}. It is important to note that in this paper depinning of a single dislocation line is considered, and that this situation is conceptually not related to that of the dynamics of complex interacting dislocation ensembles.

\section*{Methods}

\subsection*{Molecular dynamics simulations}

We simulated glide of edge dislocations in random $\rm Fe_{0.70}Ni_{0.11}Cr_{0.19}$ alloys, similar in composition to 300-series stainless steels, using molecular dynamics with methods identical to those of Sills et al~\cite{sillsLineLengthDependentDislocationMobilities2020}.
An FCC lattice was utilized (i.e., we simulated austenitic stainless steel) with the Zhou et al.~embedded atom method potential~\cite{zhouFeNiCrEmbeddedAtom2018}.
For the purposes of this study, the chosen alloy system is a model ternary random solid solution.
A free standing film geometry was used with periodic boundary conditions in the $x$ and $y$ directions and free surface boundary conditions in the $z$ direction.
The orthogonal box was oriented with $[\bar{1}\,\bar{1}\,2]$ in the $x$ direction,  $[1\,\bar{1}\,0]$ in the $y$ direction, and  $[1\,1\,1]$ in the $z$ direction.
The respective dimensions of the system were $L_x=3748\,\rm\AA$, $L_y=152\,\rm\AA$, and $L_z=75\,\rm\AA$.
The dislocation glide plane was centered in the film with its normal in the $z$ direction.
A single edge dislocation with line direction $[\bar{1}\,\bar{1}\,2]$ was inserted by removing a sheet of atoms in the $(1\,\bar{1}\,0)$ plane with thickness $b=\sqrt{2}a/2$, where $a$ is the lattice parameter, and then relaxing the system.
A time step size of 0.004~ps was used.

Shear stress was applied to the system by applying opposing forces on the free surface atoms at the top and bottom of the film.
When computing the dislocation velocity, the mean glide distance traveled by the dislocation, $d$, was computed as $d = \,\overline{\Delta y}L_y /b$, where $\overline{\Delta y}$ is the mean atomic displacement in the $y$ direction of atoms above the glide plane relative to atoms below the glide plane.
The velocity was then obtained from the slope of the $d$ versus simulation time data.
For simulations where complete pinning was not observed (i.e., the dislocation moved continuously in time), each simulation had a duration of 8~ns and one simulation was analyzed at each condition (stress and temperature) using line configurations at extracted at 0.16~ns intervals (see below).
For simulations where complete pinning was observed (i.e., the dislocation became completely arrested), each simulation had a duration of 0.4~ns---which was sufficient to ensure that the dislocation reached a completely pinned state---and 10 different random alloys were used at each condition; only the final completely pinned configurations were analyzed.

To analyze the line roughness, the dislocation extraction algorithm (DXA)~\cite{stukowskiExtractingDislocationsNondislocation2010} was utilized to extract dislocation line profiles using OVITO~\cite{stukowskiVisualizationAnalysisAtomistic2010}. 
Given that the dislocations in FCC metals disassociate into a pair of Shockley partial dislocations, line extraction via DXA produced two partial dislocation line profiles for every simulation snapshot which were analyzed independently (i.e., every MD snapshot produced two line profiles).
Part of DXA is a line coarsening and smoothing step, which removes roughness in the dislocation line resulting from the lattice triangulation that is employed by the method.
Given that our primary focus here is on the roughness of the lines, there is a concern that the smoothing algorithm in DXA may introduce artifacts into the line geometries.
In Appendix A, we discuss the possible influence of smoothing on our results and findings.
For all results presented in the body of the manuscript, we used the default coarsening and smoothing parameters in OVITO ({\tt line\_point\_separation}~=~2.5, {\tt line\_smoothing\_level}~=~1).
Based on the consistent trends and limiting behaviors of our Hurst exponents and correlation lengths, we do not believe that line smoothing affects any of our major findings.  

\subsection*{Determination of steady state}

\begin{figure}[h!]
  \caption{\csentence{Determination of the steady state velocity of the dislocation line.}
      A representative example of the dislocation velocity ($v$) as a function of time ($t$) at $T=200$~K and $\sigma=65$~MPa. The initially straight dislocation becomes rough in the initial transient regime and reaches a steady state at time $t_{\rm ss}$. From that time the average steady state velocity $v_{\rm ss}$ remains constant. \label{fig:steady_state_determination}}
  \includegraphics[width=0.6\linewidth]{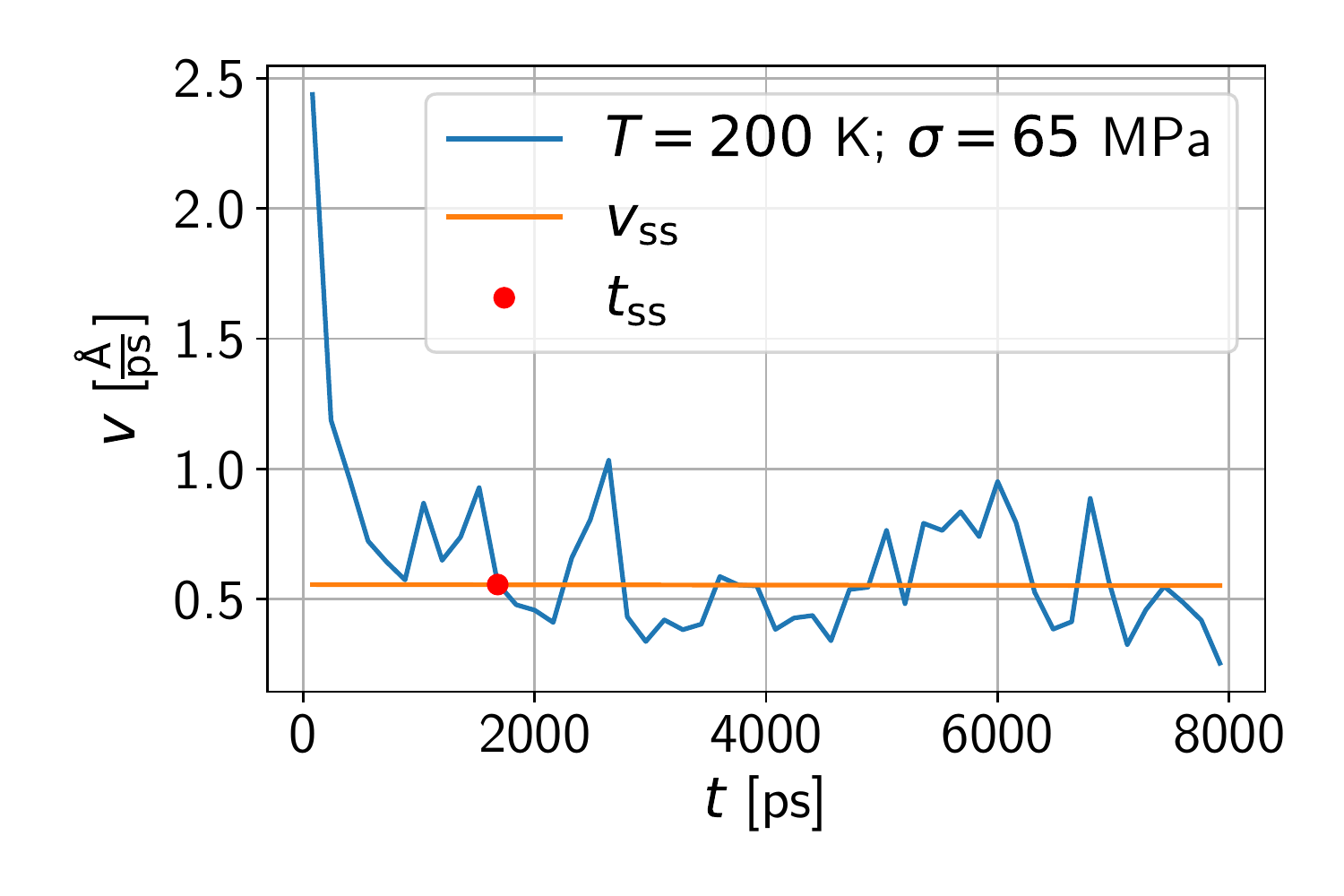}
\end{figure}

As it was mentioned in the previous section, in the beginning of the simulations the dislocation was straight. In the field of the random pinning forces exerted by the solute atoms, this shape is highly unstable. As a result the line roughens until it reaches a steady state which is either pinned for low external load or mobile for larger loads. In order to determine the steady state roughness (which is the focus here) the initial relaxation transient needs to be eliminated. 
Specifically, we must determine the time $t_\text{ss}$ at which the system reaches its steady state velocity $v_\text{ss}$ (and we assume that the line roughness has also reached steady state).
To this end, the following procedure was implemented. First the velocity $v$ of the dislocation was computed at time $t$, see Fig.~\ref{fig:steady_state_determination} for a representative example. Then the test function $f(t) = a_0 t + v_0$ was fitted to the domain $[t_0, t_\text{end}]$, where $t_0$ is an arbitrary time point and $t_\text{end}$ is the duration of the simulation. The value of $t_0$ was gradually increased from $t_0=0$ until the value of the acceleration $a_0$ dropped below a pre-defined threshold (being $8\times10^{-6}\,\rm\AA/ps^2$ or smaller). 
After this fit is performed, we take $t_\text{ss}=t_0$ and $v_\text{ss}=v_0$.
The outcome of the method is demonstrated in Fig.~\ref{fig:steady_state_determination}. The steady state is, thus, defined based solely on the velocity of the dislocation. It was checked, however, that parameters describing the roughness (correlation length and Hurst exponents) also reached a steady value after $t_{ss}$.





\subsection*{Roughness analysis}

\begin{figure}[h!]
    \caption{
    \csentence{Roughness of the dislocation lines at $T=5$ K and various stress levels below the depinning transition.}
    Roughness develops a power law dependence up to the correlation length $\xi$ (denoted by dots) characterized by the Hurst exponent $H_1$, which gives the slope of the curve. Upon increasing stress the Hurst exponent $H_1$, as well as the correlation length, increases. Above the correlation length, a different exponent $H_2$ is obeyed. Dashed lines show values of $H_1$ and $H_2$ and dots show values of $\xi$, all three obtained by the Higuchi method.
    \label{fig:hurst-compare}}
  \includegraphics[width=0.95\linewidth]{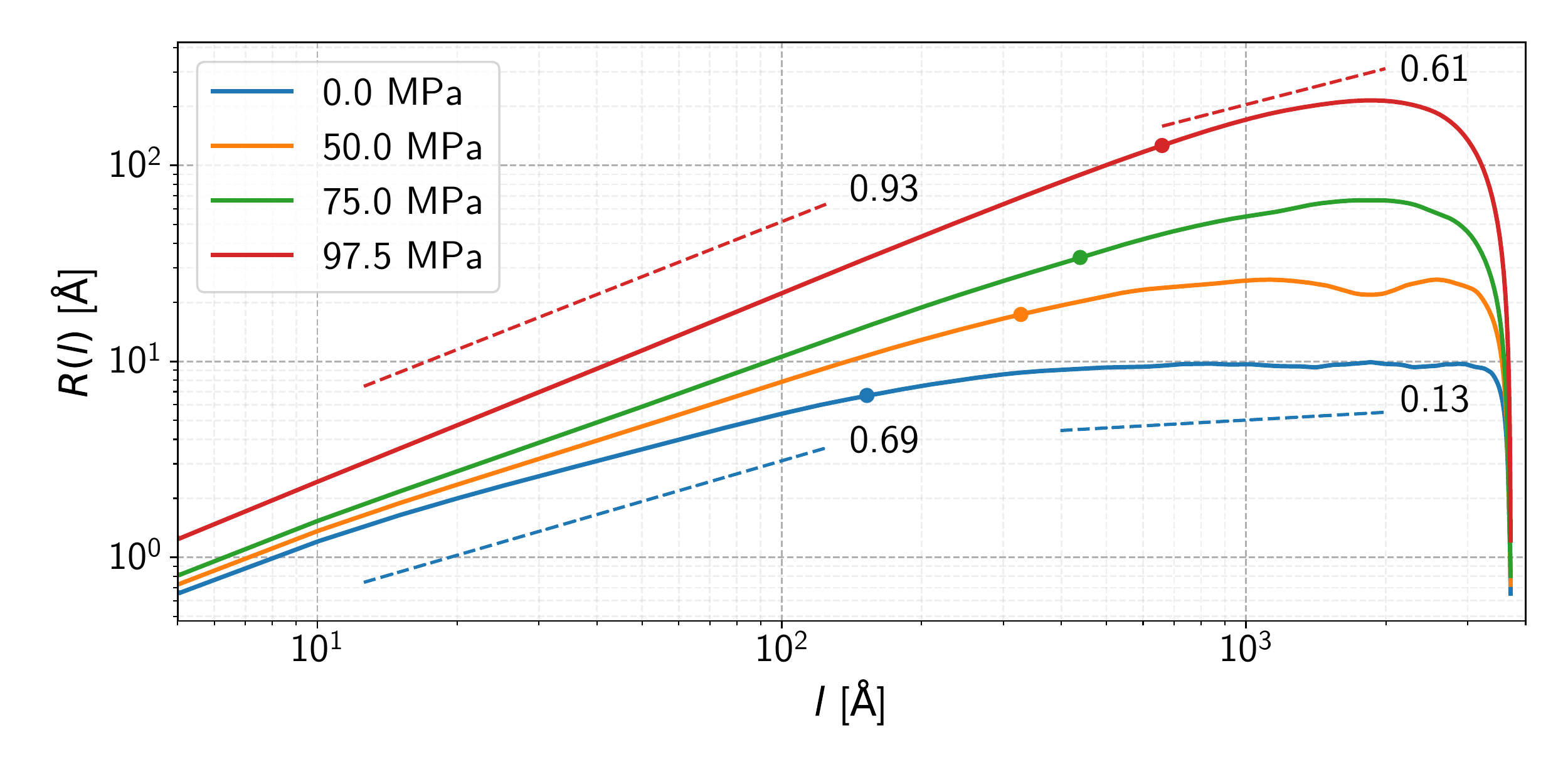}
\end{figure}

In order to describe the shape of dislocation lines, the roughness function $R(l)$ of Eq.~(\ref{eq:roughness}) was determined for dislocation profiles extracted by DXA. Figure \ref{fig:hurst-compare} shows the $R(l)$ functions for $T=5$ K at different applied load levels. As expected (see Eq.~(\ref{eq:hurst})), the plots clearly exhibit a power law regime with Hurst exponent $H_1$ up to a length scale (correlation length) above which a different power law with Hurst exponent $H_2$ is obeyed. The correlation length as well as the power law exponents increase as the stress is increased.

%

Several methods have been formulated to extract the Hurst exponents and the correlation length from data describing a rough surface \cite{krakovsk2016fractal}. In this paper we utilize the Higuchi method \cite{HIGUCHI1988277} which is a very efficient method to calculate the fractal dimension (and, thus, the Hurst exponent) of data series \cite{LIEHR2020132265}. In addition, an outstanding property of the method is that it is particularly suited for problems with two scaling regimes characterized by two different Hurst exponents. These properties explain its widespread usage within different fields of science \cite{ahammer2011, GalvezCoyt2012, Wajnsztejn2016HiguchiFD, pmid30246789, AlNuami2018}. In the following the concepts of this method are briefly introduced.

To apply the Higuchi method, an ordered data series is needed where the datapoints follow each other in equidistant steps as a function of, e.g., time or, as in our case, distance along the $x$ axis. The notation $y_i = y(i h)$ is introduced, where $h$ denotes the spacing between neighboring data points in the $x$ direction and $1 \le i \le N$ with $N$ being the total number of data points.
DXA does not output dislocation line geometries with a uniform nodal spacing, however, so we re-sampled (interpolated) the DXA output with a grid spacing of $h\approx5\,\,\rm\AA$ (defined $h$ based on the first line segment on the line).

\begin{figure}[h!]
  \caption{\csentence{Re-sampling in the Higuchi method}
      The data on an equidistant grid, shown as brown dots, are re-sampled in order to obtain $L_{k}(m)$. The green dots mark the re-sampled datapoints $y_{m+ik}$ used for the evaluation for a given $k$ and $m$. The orange square symbolises the first datapoint in the re-sampled set. \label{fig:present_higuchi}}
  \includegraphics[width=0.9\linewidth]{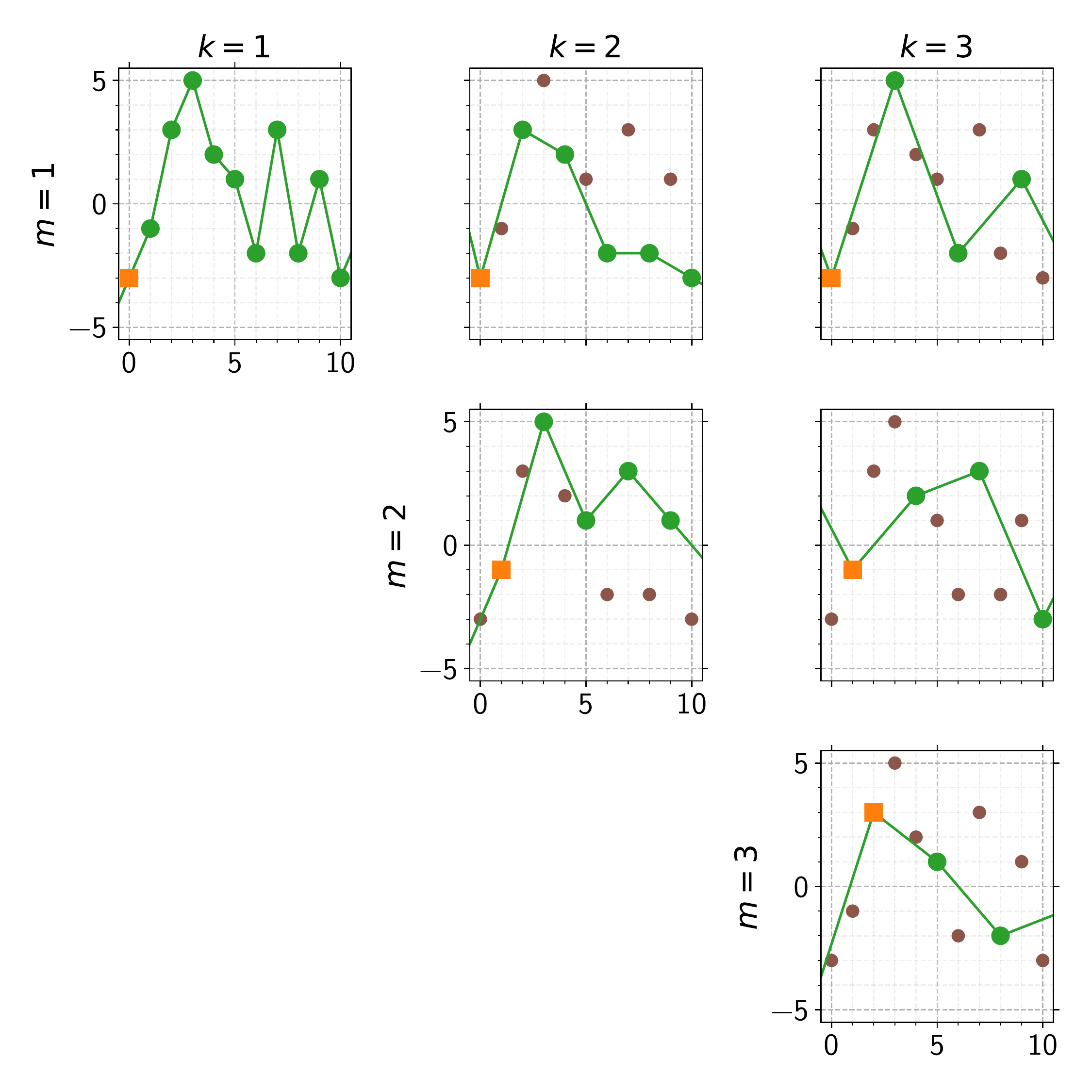}
\end{figure}

The basic idea of the method is to re-sample the data $y_i$ with different grid spacings $kh$, where $k>0$ is an integer, and then compute the roughness at the grid spacing, i.e., with $l=kh$.
The re-sampling procedure is shown in Fig.~\ref{fig:present_higuchi}.
$k$ is like a ``zoom'' factor which selects the scale at which the data is analyzed.
The starting data point for the re-sampling is denoted by the integer $m$, with $0\le m<k$.
In the Higuchi method, the following scaled roughness function is then computed for each value of $k$ and $m$:
\begin{align}
    L_k(m)=\frac{N-1}{k^2}\left(\frac{1}{\left[\frac{N-m}{k}\right]} \sum\limits_{i=1}^{\left[\frac{N-m}{k}\right]} \left\vert y_{m+ik}-y_{m+(i-1)k}\right\vert\right).
\end{align}
Note that since $(N-m)/k$ is the number of data points in the re-sampled dataset, the quantity in parentheses is similar to the average roughness when $l=kh$.
For every magnification level $k$, these $L_k(m)$ lengths are then averaged for every $m$ as $L_k = \langle L_k(m) \rangle$. Averaging over all $m$ values ensures that all data points are re-sampled equally. If the line is a fractal with dimension $D$, it is observed that $L_k \propto k^{-D}$. Therefore, on the so-called Higuchi plot, where $L_k$ is plotted as a function of $k^{-1}$ on double logarithmic axes,  the slope corresponds to the fractal dimension $D$. The Hurst exponent then follows as $H=2-D$\footnote{Accordingly, note that $L_k k^2\propto k^H$.}.


\begin{figure}[h!]
    \caption{\csentence{Higuchi method applied at $T=5$ K and $\sigma=0$ MPa.}
    Total scaled roughness $L_k$ as a function of inverse magnification $\frac{1}{k}$ (heavy blue line). Two power laws are fitted to the small (orange) and to the large (green) length scales separated by the optimal $k_\text{s}$ (marked with a dashed vertical black line). The correlation length $\xi$ is determined from the intersection point of the two power law fits (marked with a dashed vertical red line) and the Hurst exponents $H_1$ and $H_2$ from the corresponding slopes. The inset shows the fit score function $S(k_\text{s})$ as a function of $\frac{1}{k_\text{s}}$. The red dot marks the minimum value that yields the optimal $k_\text{s}$.
    \label{fig:higuchi-applied}}
  \includegraphics[width=0.95\linewidth]{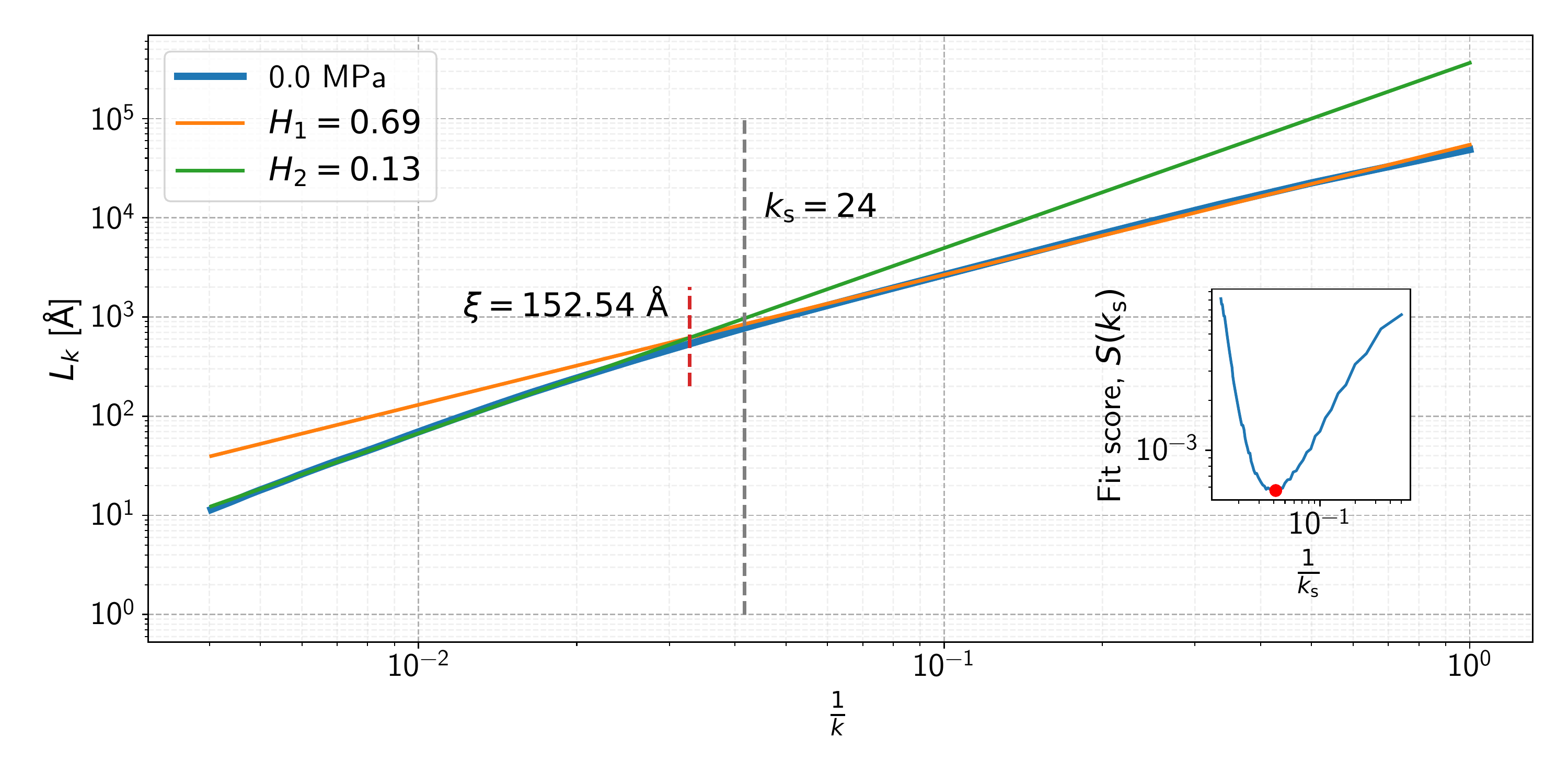}
\end{figure}

If there are two scaling regimes separated by a correlation length, then a crossover appears at a certain $k_\text{s}^{-1}$ value, as seen in Fig.~\ref{fig:higuchi-applied} (see below for how $k_s$ is obtained). 
In all cases studied in this paper we indeed observed two scaling regimes on the Higuchi plots, so, we developed an automated fitting procedure in order to obtain the Hurst exponents and the correlation lengths without any systematic biases that could be introduced by manual fitting. In the following this method is briefly introduced.

Given the finite periodic length of the dislocation line set by the box length $L_x=3748\,\rm\AA$, re-sampling could not be accomplished for arbitrarily large values of $k$.
In practice we set $k_{\rm max} h = 1250 \, \rm\AA{}$ as an upper limit, which gives $k_{\rm max} = 250$. On the lower end $k_\text{min}= 1$ was chosen, which corresponds to all data points being used.

First, we split the domain of the Higuchi plot into two parts by introducing the variable $k_\text{s}$. Then, both the $[1/k_\text{max}, 1/k_\text{s}]$ and the $[1/k_\text{s}, 1/k_\text{min}]$ regimes were fitted with a test function of the form $A (1/k)^D$ (i.e., a pair of lines in log-log space) yielding values $(A_1, D_1)$ and $(A_2, D_2)$ for the two regimes. To estimate the quality of the fit for that particular $k_{\rm s}$ value, the score function
\begin{equation}
\begin{split}
    S(k_s) =& \frac{1}{\log \left( k_\text{max} / k_\text{s} \right)}  
    \int\limits_{1/k_\text{max}}^{1/k_\text{s}} \left[ \log(L_k) - \log(A_1 k^{-D_1}) \right]^2 \, \mathrm d \left[ \log\left( \frac 1k \right) \right] \\
    & + \frac{1}{\log \left( k_\text{s} / k_\text{min} \right)} \int\limits_{1/k_\text{s}}^{1/k_\text{min}} \left[ \log(L_k) - \log(A_2 k^{-D_2}) \right]^2 \, \mathrm d \left[ \log\left( \frac 1k \right) \right]
\end{split}
\end{equation}
was introduced which measures the squared distance between the original data and the fit functions on the log-log plot normalized with the extension of the two regimes. The optimal value of $k_\text{s}$ was determined from the global minimum of the score function evaluated numerically as seen in the inset of Fig.~\ref{fig:higuchi-applied}. Given the curvature in the roughness data near $k_\text{s}$, the intersection of the two fitted power laws is not necessarily at $1/k_\text{s}$ (as in the example of Fig.~\ref{fig:higuchi-applied}).
The correlation length $\xi$ was, therefore, computed from the intersection point of the power law fits rather than $k_\text{s}$. As seen in Fig.~\ref{fig:hurst-compare} the Hurst exponents obtained from the Higuchi method, shown as dashed lines, are fully compatible with the roughness plots, as expected.

\section*{Results}
\subsection*{Depinning of dislocation lines}
\begin{figure}[h!]
  \caption{\csentence{Steady state dislocation velocity $v_{\rm ss}$ as a function of temperature $T$ and external stress $\sigma$.}
      The data obtained from the MD simulations were fitted with the form of Eq.~(\ref{eq:v_depinning}) yielding parameter values shown in the figure legend.
      \label{fig:speed-stress-T}}
  \includegraphics[width=0.8\linewidth]{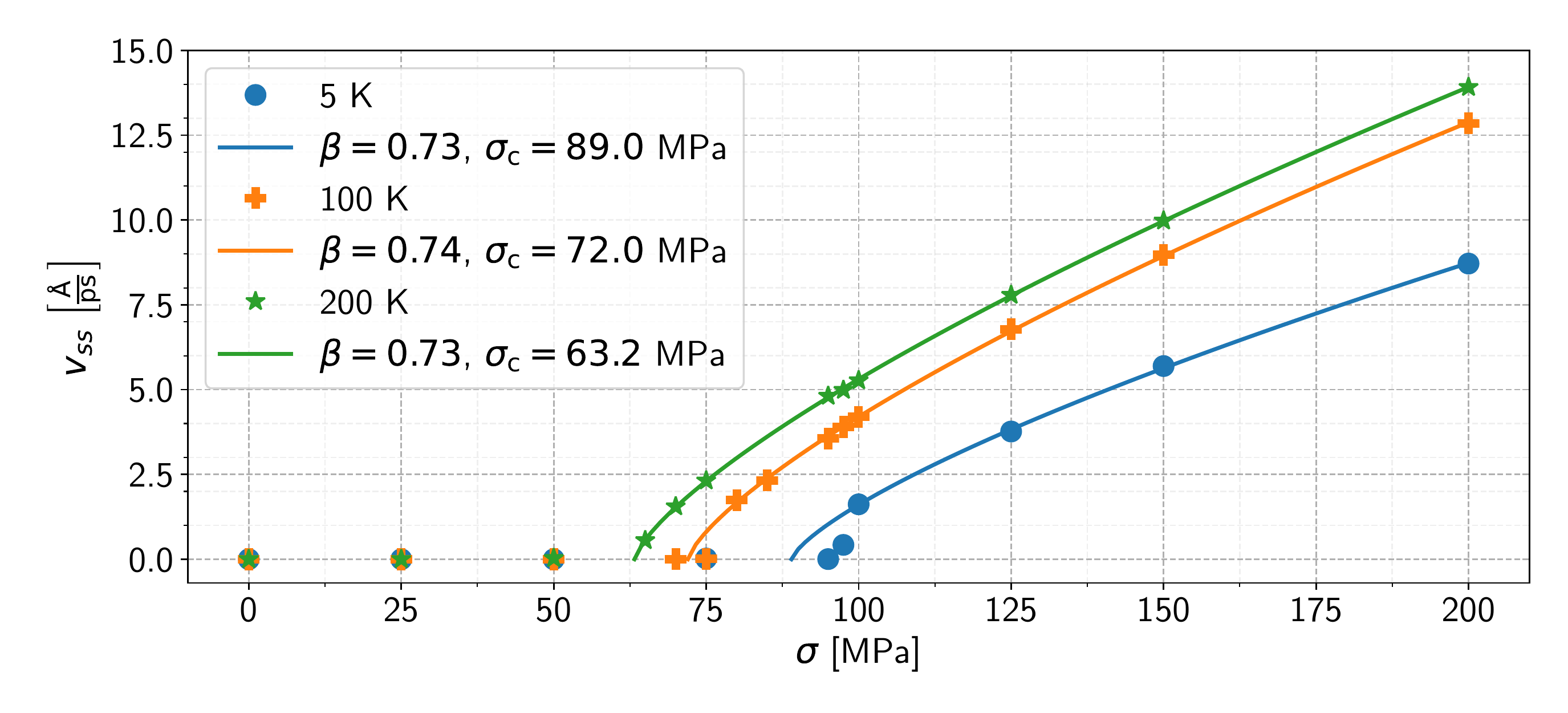}
\end{figure}

In order to determine the depinning threshold of dislocations (i.e., the critical or friction stress), the steady state velocity $v_{\rm ss}$ was determined as a function of the applied stress $\sigma$. According to Fig.~\ref{fig:speed-stress-T}, the observed behavior is analogous to a depinning transition: dislocations are pinned below a critical stress $\sigma_\text{c}$ and move continuously above. In the mobile regime the velocity obeys $v_\text{ss} \propto(\sigma- \sigma_\text{c})^\beta$ with $\beta\approx0.73$, close to the value of $\beta=0.62$ obtained from discrete dislocation dynamics simulations of impenetrable precipitates~\cite{bako2008dislocation}. Thermal rounding is also observed in the gradual decrease of $\sigma_\text{c}$ as the temperature increases from 5 K to 200 K. At these temperatures, however, thermal creep was not observed below $\sigma_\text{c}$.
Considerable thermal creep was observed at higher temperatures in our prior work~\cite{sillsLineLengthDependentDislocationMobilities2020}.
For reference, we observed complete pinning ($v_\text{ss}=0$) under the following conditions: $T=5$~K and $\sigma=0$, 25, 50, 75, 95~MPa; $T=100$~K and $\sigma=0$, 25, 50, 75 MPa; and $T=200$~K and $\sigma=0$, 25, 50 MPa.
We note that we observed complete pinning at $T=5$~K when $\sigma=95$~MPa despite the fact that this stress is above the depinning stress of $\sigma_\text{c}=89$~MPa obtained from the fit in Fig.~\ref{fig:speed-stress-T}.
We believe that this is an artifact of the finite box size since the correlation length $\xi$ becomes very large near the depinning transition (see next section).


\subsection*{Roughness of dislocation lines}

\begin{figure}[h]
    \caption{\csentence{Dislocation line geometries at $T= 5\,\text{K}$ and different stress levels.}
    Three exemplary dislocation lines for each case as extracted by DXA (shifted so that their average $y$ coordinate was set to zero).
    The magnified parts are 500 \AA{} wide and 80 \AA{} high.
    \label{fig:dislocation-lines}}
  \includegraphics[width=0.95\linewidth]{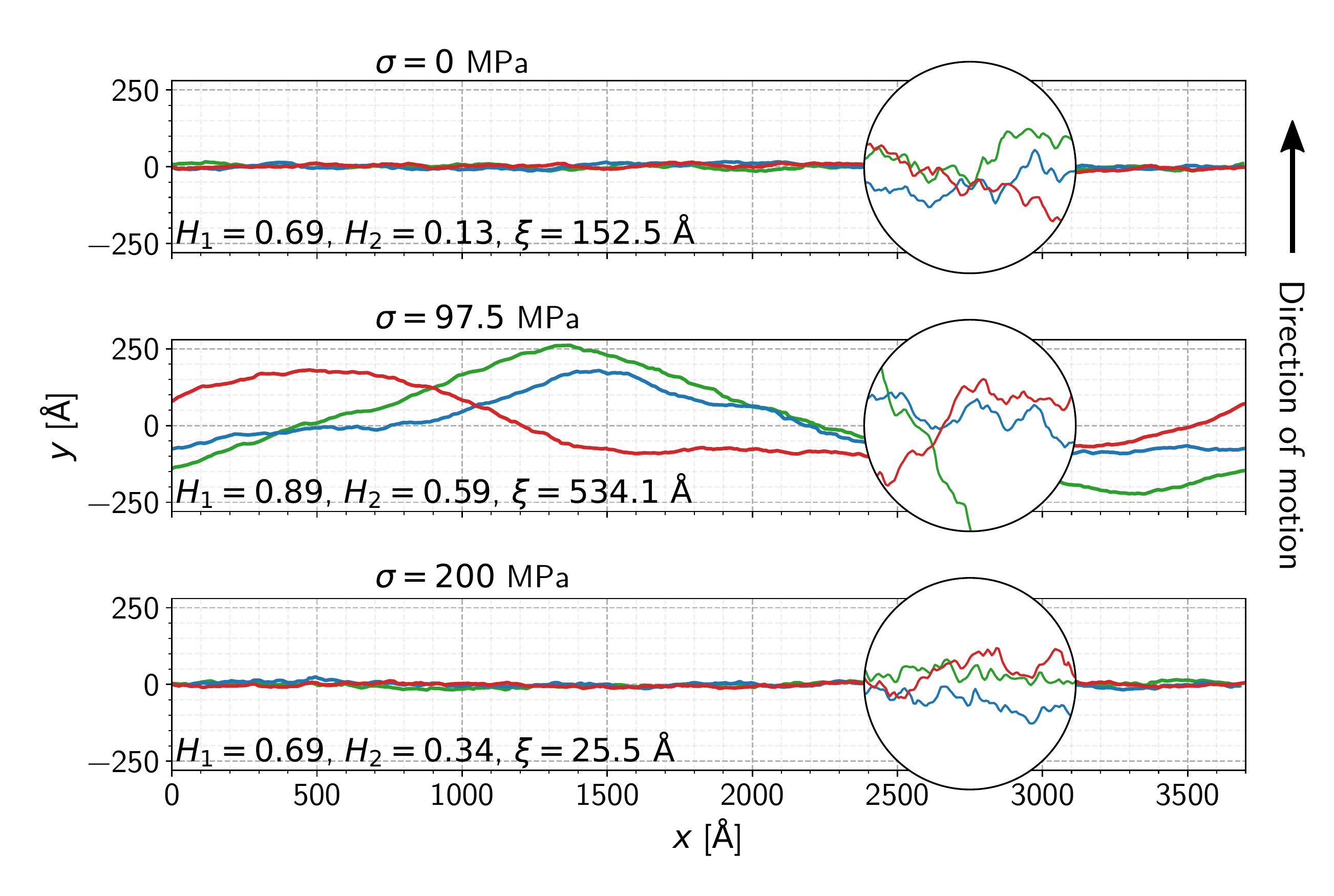}
\end{figure}

As the stress increases the shape of the dislocation lines gradually changes. Example snapshots can be seen in Fig.~\ref{fig:dislocation-lines} where representative dislocation lines are shown at $T=5\,\text{K}$. Close to depinning ($\sigma=97.5$~MPa) the morphology changes dramatically, with the dislocation bowing out significantly in the $y$ direction over large length scales. In contrast, away from the depinning transition ($\sigma=0$, $\sigma=200$~MPa) the line remains essentially straight over large length scales.
Furthermore, at small scales, shown in insets, fluctuations at short wavelengths (i.e., small $l$ values) seem more pronounced away from the depinning transition.


\begin{figure}[h!]
\caption{\csentence{Hurst exponents and correlation lengths of the dislocation lines for various temperatures $T$ and external stress levels $\sigma$.}
    (Left) Hurst exponents $H_1$ and $H_2$ determined with the Higuchi method as a function of the applied stress $\sigma$. $H_1$ corresponds to small $k$ (and $l$) values and $H_2$ corresponds to large $k$ (and $l$) values. 
    (Right) Correlation length values $\xi$ at which the Hurst exponent changes from $H_1$ to $H_2$ as a function of $\sigma$. 
    The purple region shows the stress range within which the depinning transition is expected based on our simulation data. 
    The vertical black dotted lines show the critical stress $\sigma_\text{c}$ according to Fig.~\ref{fig:speed-stress-T}. 
    The gray area shows where the smoothing of the DXA algorithm could alter the result.\label{fig:stress-t-d-xi}}
  \includegraphics[width=0.95\linewidth]{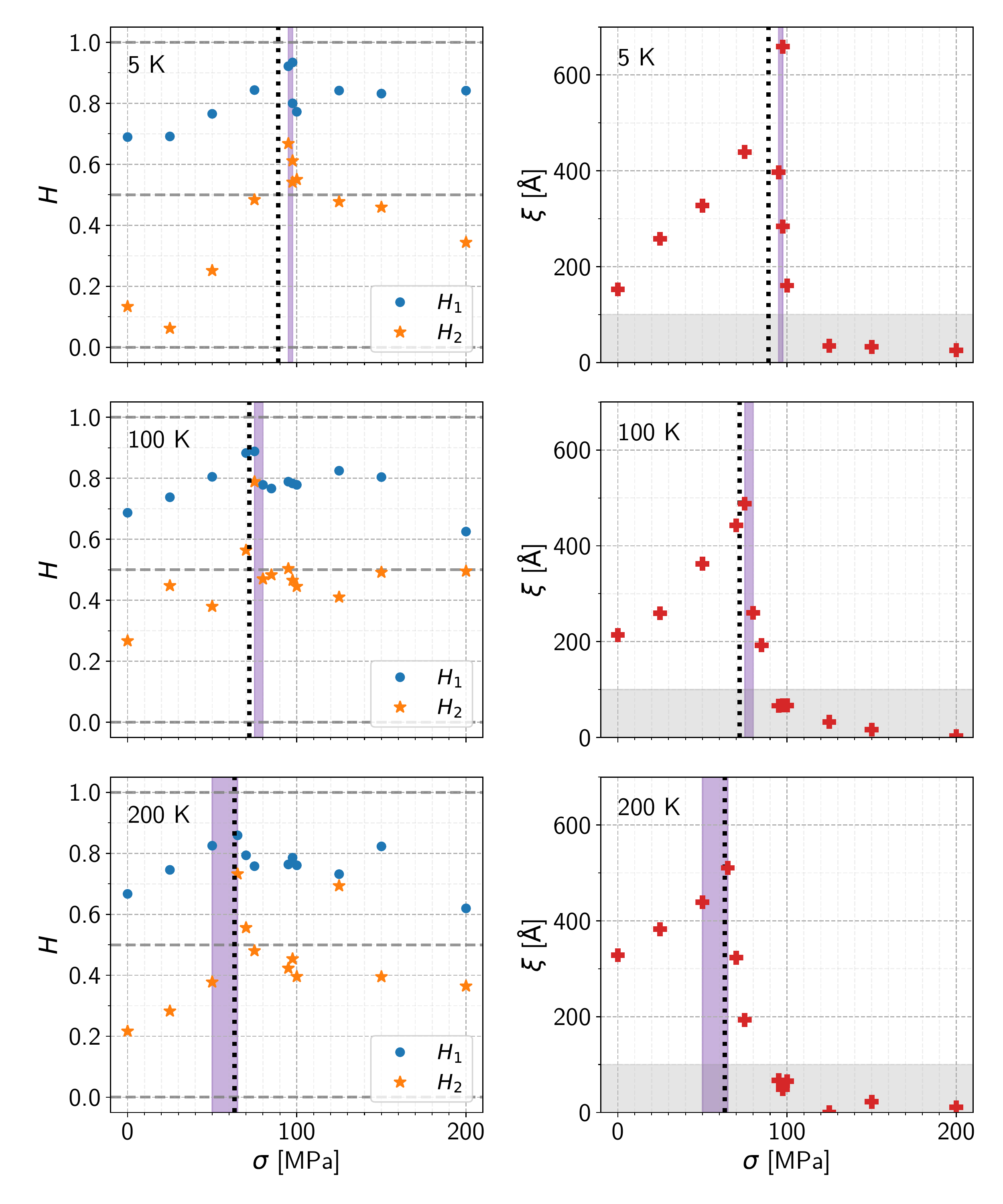}
\end{figure}

These observations can be quantified using the Higuchi method introduced above. According to the analysis, two scaling regimes can be observed which are characterized by Hurst exponents $H_1$ and $H_2$ and separated by the correlation length $\xi$. Figure \ref{fig:stress-t-d-xi} summarises the values obtained by the automated fitting procedure; the results contained in Figure \ref{fig:stress-t-d-xi} are the principal contribution of this work.

The behavior of the correlation length $\xi$ is typical to that of a depinning transition: it increases as $\sigma_\text{c}$ is approached and decreases above. At $\sigma_\text{c}$ the divergence predicted by Eq.~(\ref{eq:xi}) is not observed since in this region $\xi$ becomes comparable to the system size $L_x=3748\,\text{\AA}$. Therefore, larger simulation volumes would be required to determine the value of the exponent $\nu$. The effect of temperature on $\xi$, apart from the shift in the depinning threshold discussed above, is weak; due to thermal noise, correlation lengths are slightly larger at higher temperatures when $\sigma<75$~MPa (far from the depinning transition).

The Hurst exponents $H_1$ and $H_2$ below and above the correlation length, respectively, show a clear trend that is again only weakly affected by increasing temperature. Namely, in the pinned state the small scale exponent $H_1$ increases from $H_1 \approx 0.7$ at $\sigma = 0$ towards $H_1\approx 1.0$ at $\sigma = \sigma_\text{c}$ whereas the large scale exponent is significantly smaller, typically $0 \le H_2 \le 0.4$. This means that at small stresses the roughness at small length scales is almost random due to thermal noise with $H_1\approx0.5$ (even at $T=5\,\text{K}$)~\cite{geslin2018thermal, zhai2019properties} and with increasing stress its shape gradually approaches that found for quenched disorder with $H_1\approx1$~\cite{zapperi2001depinning, hiratani2003dislocation, bako2008dislocation, zhai2019properties,patinetAtomicscaleAvalancheDislocation2011}. Such gradually changing exponents are not commonly observed in depinning transitions and the possible reason and implications will be discussed in the Discussion section below.

Above the depinning threshold the correlation length $\xi$ quickly drops below 100~\AA~and becomes comparable to the dislocation core radius (separation distance between the partial dislocations). Above this scale (i.e., $l>\xi$) the dislocation line develops scale-free roughness with Hurst exponent $H_2 \approx 0.5$. This is equivalent to the case of thermal noise \cite{geslin2018thermal, zhai2019properties}.

\section*{Discussion}

We begin the discussion by formulating a qualitative picture that explains the trends observed above, obtaining a description of the depinning process that is very similar to that of Patinet et al.~\cite{patinetAtomicscaleAvalancheDislocation2011}.
First, we focus on the behavior prior to the depinning transition ($\sigma<\sigma_\text{c}$).
Consider a periodic dislocation line of length $L$ under applied shear stress $\sigma$ and temperature $T$. 
At every position along the dislocation, the line may glide forward under the applied stress.
Since the solid solution contains many pinning points (solute atoms), the line will tend to locally bow out between a set of ``dominant'' pinning points.
We refer to the distance between these dominant pinning points as the \emph{bowing length}, $l_\text{b}$, and the distance that the line bows out as the bowing amplitude.
For dilute solid solutions (strong pinning), this interpretation is exact since solutes are encountered one-by-one.
With a dense solid solution (weak pinning) as analyzed here, the story is more complicated given that the solutes interact with the dislocation line over the entire bowing length.
None-the-less, we believe that the basic picture still holds in both cases.

At every point along the line, the roughness is controlled by the balance between the line tension which acts to straighten the line, the applied stress which acts to bow out the line, and the solute forces which act to induce a random (but correlated) line geometry. 
When the bowing length is small the dislocation line interacts with relatively few solutes, making it easier for the line to bow out coherently since the probability of encountering a ``strong'' interaction is lower.
As a result, the line is able to bow out at small bowing lengths, which is the source of the roughness.
This roughness (bowing amplitude) increases with bowing length, as predicted by Eq.~(\ref{eq:hurst}), because the line is able to bow out more when the bowing length is longer.
This gives rise to correlated roughness with $0.5<H_1<1$.
As the bowing length $l_\text{b}$ increases, the probability of encountering a ``strong'' interaction increases, making it increasingly difficult for coherent bow outs of length $l_\text{b}$ to form; the strong interactions ``break up'' long bowing lengths.
Eventually, it becomes nearly impossible for such bow outs to form, causing the line roughness to become uncorrelated (or even anti-correlated) beyond a certain bowing length with $0<H_2<0.5$.
This is precisely what the correlation length $\xi$ represents: $2\xi$ is the maximum length scale at which coherent bow outs can form (the factor of two accounts for the fact that roughness is measured in the middle of a bow out).
A bow out with a length above $2\xi$ is broken into smaller bowing lengths by the solute field.
When $\sigma=\sigma_\text{c}$, the stress is high enough that the applied stress is able to overcome the solute pinning forces at all length scales, causing $\xi\rightarrow\infty$.
Hence, bow outs at any length scale are able to form at the depinning transition.

At finite temperature the picture becomes complicated by the fact that the pinning induced by solutes is no longer deterministic, since the line may overcome a set of solute obstacles with the aid of random thermal fluctuations: motion of the dislocation line becomes thermally activated.
None-the-less, the basic picture still holds, as our data show.
The correlation lengths $\xi$ and Hurst exponents $H_1$ and $H_2$ follow identical trends at $T=5$, 100, and 200~K, albeit with noisier data at higher temperatures.
Interestingly, the variation of $\xi$ with stress is shown to be temperature-dependent.
For example, at $\sigma=50$~MPa we find that $\xi=327$, 363, and 439~\AA~when $T=5$, 100, and 200~K, respectively.
This indicates that as temperature is increased, larger bow outs are able to form since thermal activation ``reduces'' the pinning strength of the solutes. 

The influence of finite temperature also explains the trends we observe in $H_1$ and $H_2$.
Recently, the work of Zhai and Zaiser~\cite{zhai2019properties} showed how, prior to the depinning transition, the Hurst exponents behaves in two limits.
In the absence of solutes at finite temperature, random thermal noise leads to Hurst exponents $H_1=H_2=0.5$.
Geslin and Rodney obtained the same result by analyzing line geometries from MD simulations of pure Al~\cite{geslin2018thermal}.
At the other limit, a dislocation interacts with a solute field at $T=0$~K giving Hurst exponents of $H_1=1$ and $H_2=0$.
Our results correspond to a mixture of these two limits, since we consider a solid solution system at finite temperature.
Hence, prior to the depinning transition we should expect that $0.5<H_1<1$ and $0<H_2<0.5$, which is exactly what is observed in Fig.~\ref{fig:stress-t-d-xi}.
At low stresses, where thermal noise strongly influences the roughness, $H_1\approx0.7$.
As the stress increases, so does $H_1$ since the thermal noise is washed out by the sampling of the solute field; the dislocation line is able to find stronger and stronger pinning sites which overwhelm the random thermal forces.
Finally at the depinning transition we find that $H_1\approx0.9$.

After the depinning transition ($\sigma>\sigma_\text{c}$), the behavior simplifies a bit because the solutes simply introduce another random noise field which the dislocation samples (as is observed during domain wall migration in ferromagnetics \cite{duemmer2005critical, bustingorry2010random}).
Accordingly, we observe that both $H_1$ and $H_2$ tend to 0.5 when $\sigma>\sigma_\text{c}$.
Furthermore, we observe that $\xi\rightarrow0$ above the depinning transition, which is also consistent with uncorrelated roughness; the line geometry converges to an uncorrelated fractal.

\subsection*{Comparisons with solid solution strengthening theories}

As mentioned in the Introduction, most theories for solid solution strengthening are formulated on the basis of length scales which govern the dislocation's interaction with the solute field. 
Dating back to the variational analysis of Mott~\cite{mottMechanicalStrengthCreep1952,ardellPrecipitationHardening} and more recently formulated by Argon~\cite{argonStrengtheningMechanismsCrystal2008}, an ad hoc interaction range $\omega$ is commonly defined, beyond which the dislocation-solute interaction is assumed to go to zero.
Using this approximation in conjunction with the statistics of random point distributions, estimates are obtained for the mean bowing length at zero applied stress, $\bar{l}_\text{b0}$. For example, Mott obtained that~\cite{mottMechanicalStrengthCreep1952}
\begin{equation}
    \bar{l}_\text{b0} = \left(\frac{2\Gamma L_s^4}{F_s \omega}\right)^{1/3}
\end{equation}
where $\Gamma$ is the dislocation line tension, $F_s$ is the strength of the solutes, and $L_s$ is the mean solute spacing in the glide plane.
Argon obtained a similar result~\cite{argonStrengtheningMechanismsCrystal2008}.
The flow stress is then directly related to the mean bowing length $\bar{l}_\text{b0}$ via a force balance or thermal activation analysis~\cite{ardellPrecipitationHardening,argonStrengtheningMechanismsCrystal2008}.
Determination of the mean bowing length at zero-stress $\bar{l}_\text{b0}$ is thus a critical feature of these classical models.

More recently, the model of Leyson et al.~\cite{leysonQuantitativePredictionSolute2010} advanced theoretical understanding of solid solution strengthening by taking an atomistic view of the problem of determining $\bar{l}_\text{b0}$.
In their theory, the total energy of the dislocation line is estimated by accounting for the line energy and the interaction energy with the solute field.
The mean bowing length and bowing amplitude is then obtained by minimizing the total energy (similar to Mott's analysis) at zero applied stress.
Again, once these length scales are determined strengthening due to solutes can be computed.
A major benefit of Leyson et al.'s theory is that there is no need to assume an ad hoc interaction length scale $\omega$; instead, solute interactions at all scales are rigorously incorporated.

\begin{figure}[h!]
    \caption{\csentence{Power spectral density at $T=5$ K, below the depinning stress.}
    As the stress tends towards the depinning stress only the small-$q$ (large wavelength) part changes, corresponding to an increase of the correlation length (i.e., to the appearance of larger bow-outs).
    \label{fig:psd}}
  \includegraphics[width=0.95\linewidth]{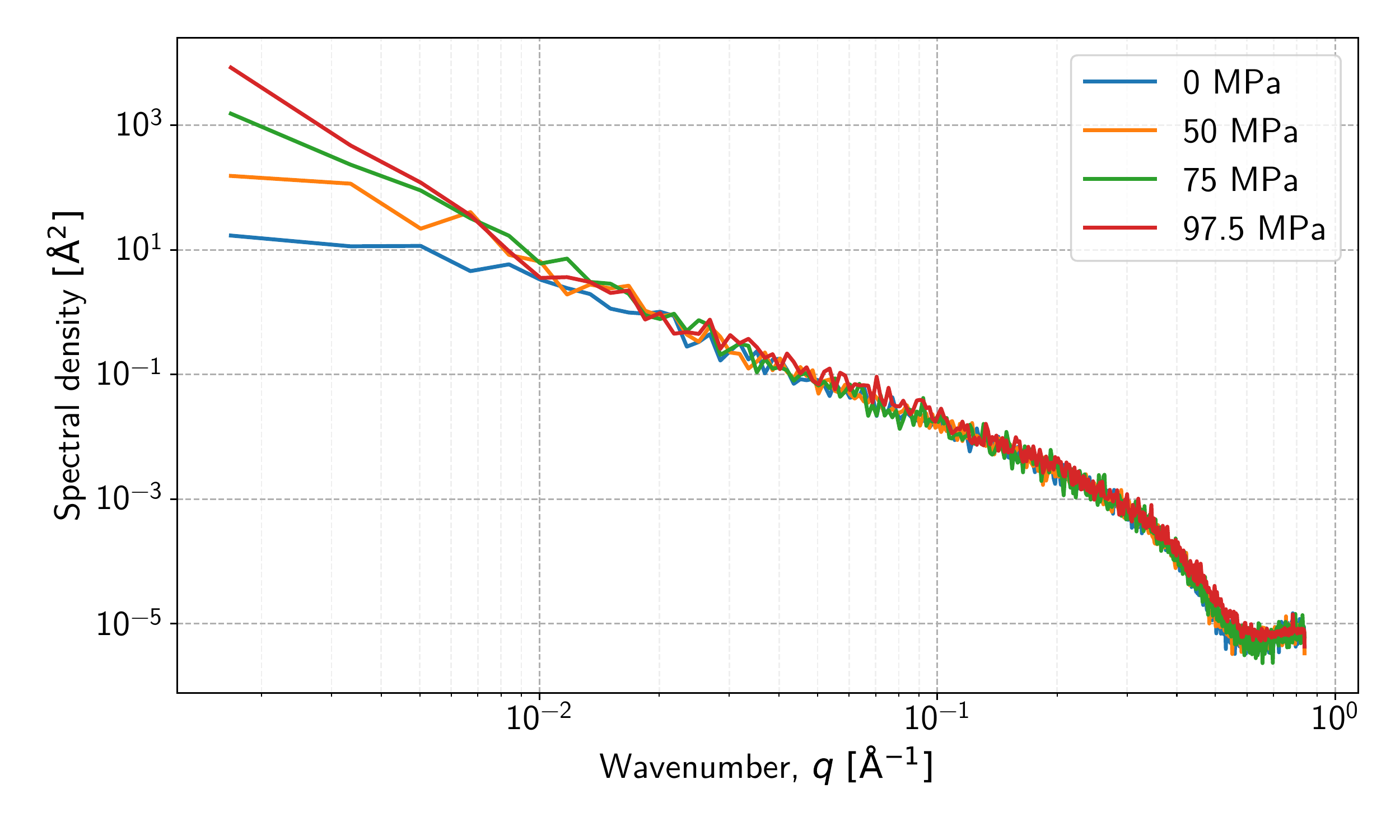}
\end{figure}

The common feature among all of these theories is that strengthening is governed by the mean bowing length at zero-stress $\bar{l}_\text{b0}$. 
As even the earliest theories acknowledged, however, the process of a dislocation line sampling the solute field is statistical leading to a distribution of bowing lengths.
Our results bolster this notion, showing that many bowing length scales operate simultaneously.
In fact, \emph{all} bowing length scales below $2\xi$ are operative.
This observation brings into question the geometric assumptions of previous solid solution strengthening theories.
For example, in Leyson et al.'s theory it is assumed that the line adopts a quasi-sinusoidal profile characterized by a single bowing length.
Our results indicate that this picture simply isn't realistic: the line is characterized by many bowing lengths, none of which are clearly dominant, 
To help demonstrate this point, we plot the average (over all relevant DXA snapshots) power spectral density of the dislocation line at $T=5$~K and $\sigma=0$, 50, 75, and 97.5~MPa in Fig.~\ref{fig:psd}.
If a single bowing length $l_\text{b}$ were dominant, we would expect to see a peak in the power spectral density at $q=2\pi/l_\text{b}$ ($q$ is the wavevector).
Clearly no such peak exists.
In a subsequent analysis, Leyson and Curtin considered the possibility of bowing at multiple length scales, concluding that at high temperatures a second bowing length scale becomes active~\cite{leysonSoluteStrengtheningHigh2016}.
Again, this finding seems inconsistent with our results.
None-the-less, it may be the case that the mean bowing length $\bar{l}_\text{b0}$ does govern strengthening, so that these observations are inconsequential.
It is unclear whether the specific form of the bowing length distribution affects solid solution strengthening or the mobility of dislocations in solid solutions.
Future research should be focused on assessing the bowing length distribution and its influence in more detail, for example using the theoretical framework outlined below.

An additional point to raise is that since $\xi$ increases with stress below the depinning transition, we expect that the mean bowing length will also increase with stress.
Hence, it is not obvious that the mean bowing length at zero-stress---as is the focus of the theories mentioned above---is the relevant length scale which dictates a dislocation's response under non-zero stresses.
Somewhat troublingly, we note that at the depinning transition, i.e., the yield stress, \emph{there is no mean bowing length} because $\xi\rightarrow\infty$ so that all bowing length scales are operable.
At this point it is unclear how consequential this fact is with respect to existing solid solution strengthening theories.

\subsection*{A mobility model for $\sigma<\sigma_\text{c}$}

We can use the above interpretation of the dislocation interaction with the solute field to develop a mobility model.
This mobility model is valid for the ``creep'' regime prior to the depinning transition $\sigma<\sigma_\text{c}$, where the dislocation interacts with the solute field in a thermally activated manner.
Our goal in formulating this model is to clarify the fundamental physics at play and how they influence the overall mobility of the dislocation line.
We also seek to better understand previous observations of length-dependent mobilities at low stresses and temperatures~\cite{sillsLineLengthDependentDislocationMobilities2020}.
In contrast to previous theories where the mean bowing length is the focus, we instead seek a more comprehensive picture where all relevant bowing lengths are considered.

We wish to answer the following question: for a given stress $\sigma$, temperature $T$, and local solute distribution $\{{\bf x}_i^s\}$ (where ${\bf x}_i^s$ is the coordinate of solute $i$ with species $s$ and brackets denote the set of all solute atoms), what is the mobility of the dislocation line at a point $x$?
The analysis above indicates that line segments of length $l_\text{b}<2\xi$ are able to bow out coherently.
When $T=0$ and $\sigma<\sigma_\text{c}$, a distribution of bow outs forms and then remains pinned in a final state of equilibrium; below the critical stress the mobility is zero.
When $T>0$, however, it becomes possible for a bow out to overcome its local solute environment via thermal activation and begin gliding in avalanche-style motion~\cite{patinetAtomicscaleAvalancheDislocation2011}.
Hence, determining the mobility of the dislocation line is akin to determining the release rate for a given bowing length $l_\text{b}$ at stress $\sigma$ and temperature $T$.
In this view, the line is constantly snapping from one metastable state to another across a range of local bowing length scales.
This view is consistent with experimental observations of domain wall migration in ferromagnetic materials \cite{durin2000scaling, grassi2018intermittent} and previous MD simulations of dislocation motion in a random alloy~\cite{patinetAtomicscaleAvalancheDislocation2011}.

We can put these statements into precise mathematical terms as follows.
First, we define a \emph{bowing density of states} for the dislocation line, $n_\text{b}(l_\text{b})$, such that $n_\text{b}(l_\text{b})dl_\text{b}$ is the number of line segments with bowing length $l_\text{b}$ per unit length of dislocation line. 
Next we invoke a stress, temperature, and bowing-length-dependent waiting duration $\bar{t}_w(l_\text{b},\sigma,T)$, which is the average time that a segment of length $l_\text{b}$ is pinned before being released by thermal fluctuations.
Finally, to combine these ingredients into a mobility model which describes the dislocation velocity, we must more precisely define what is meant by velocity.
The ``true'' dislocation velocity is of course well-defined as the speed at which the line moves locally at every point along its length.
In the picture of solute interactions here, the local velocity varies rapidly along the line and in time.
Hence, it is not a tremendously useful quantity from the standpoint of understanding plasticity.
Instead, we are often interested in an average velocity over some length scale.
For example, in discrete dislocation dynamics velocities are computed for line segments of finite length~\cite{sillsFundamentalsDislocationDynamics2016}.
We must, then, clarify a coarse-grained velocity definition to specify how the averaging is accomplished.
One option is to coarse-grain in terms of the area sweep rate of the dislocation line, $\dot{A}$, since the plastic strain rate is proportional to $\dot{A}$~\cite{argonStrengtheningMechanismsCrystal2008}.
Using this approach, we can employ the Orowan equation to define the average line velocity as
\begin{equation}
    \bar{v} = \frac{\dot{A}}{L}.
    \label{eq:vbardef}
\end{equation}
Hence, in this definition the average velocity is the area sweep rate per unit length of the dislocation line.
The area sweep rate for the dislocation line described by density of states $n_b(l_\text{b})$ can be stated as
\begin{equation}
    \dot{A} = L\int_{l_{\rm min}}^{2\xi} \left(\frac{\overline{\Delta A}(l_\text{b},\sigma,T)}{\bar{t}_w(l_\text{b},\sigma,T)}\right) n_b(l_\text{b})dl_\text{b}
    \label{eq:Adot}
\end{equation}
where $\overline{\Delta A}(l_\text{b},\sigma,T)$ is the mean increment of swept area that results from release of a segment of bowing length $l_\text{b}$ at stress $\sigma$ and temperature $T$.
$\overline{\Delta A}(l_\text{b},\sigma,T)$ can be thought of as the average avalanche magnitude associated with depinning of a bowing segment of length $l_\text{b}$.
This integral is simply a sum over the swept area contributions from all bowing lengths.
The upper limit of integration $2\xi$ is the length above which bow outs cannot form.
The lower limit is a minimum segment length set by the discreteness of the crystal lattice ($l_{\rm min}\sim b$)
\footnote{Eq.~(\ref{eq:Adot}) has an interesting analogy to models of specific heat in solids~\cite{kettersonPhysicsSolids2016}. 
In the famous Debye model, the specific heat is determined by summing over contributions from all active phonons. 
The phonon spectrum is characterized by a density of states, just like the bowing length here, giving rise to an integral of the same form as Eq.~(\ref{eq:Adot}).
The upper limit of the integral is the Debye temperature, given by the lowest frequency phonon that is operable in the same way the $2\xi$ is the largest operable bowing length here.
The lower limit is set by the discreteness of the lattice.
Finally, we note that in contrast to our mobility model which considers all bowing lengths, traditional models consider only one bowing length ($\bar{l}_0$).
This is analogous to the difference between the Debye and Einstein models for specific heat; Einstein simplified the analysis by assuming that all phonons could be represented by a single frequency (rather than a spectrum).
The Einstein model predicts an incorrect temperature-dependence of the specific heat at low temperatures, whereas the Debye theory correctly captures it ($\propto T^3$).
It remains to be seen if this feature of the analogy carries through into the analysis here.}.
Combining Eqs.~(\ref{eq:vbardef}-\ref{eq:Adot}) leads to the following mobility model:
\begin{equation}
    \bar{v} = \int_{l_{\rm min}}^{2\xi} \left(\frac{\overline{\Delta A}(l_\text{b},\sigma,T)}{\bar{t}_w(l_\text{b},\sigma,T)}\right) n_b(l_\text{b})dl_\text{b}.
    \label{eq:vbar}
\end{equation}

In principle, Eq.~(\ref{eq:vbar}) provides a route for constructing mobility laws derived from basic theoretical concepts.
Namely, the bowing density of states $n_b(l_\text{b})$, the mean increment of swept area $\overline{\Delta A}(l_\text{b})$, and the mean waiting time $\bar{t}_w(l_\text{b})$ are fundamental descriptors of how the dislocation interacts with the solute field and could be derived from atomistic or discrete dislocation models.
Our results above give some clues as to what these functions may look like.
For example, one possibility is that $\overline{\Delta A}\propto l_\text{b}^{H_1+1}$, since the amount of swept area should scale with the product of the bowing length and the roughness (bowing amplitude) of the bow out~\cite{patinetAtomicscaleAvalancheDislocation2011}.
Unfortunately, these functions are likely to depend on bowing length, stress, temperature, and the solute distribution in a complex manner.
Patinet et al.~\cite{patinetAtomicscaleAvalancheDislocation2011} demonstrated one possible technique for obtaining these parameters, considering the avalanche magnitude (swept area) and duration (waiting time) statistics associated with dislocation motion in a random alloy.
Hence, in order to make the mobility model Eq.~(\ref{eq:vbar}) useful, additional research is necessary on the fundamentals of dislocation avalanche statistics in solid solutions.
None-the-less, the analysis above gives an explicit account of what governs dislocation mobility in the creep regime.

\subsection*{Interpretation of length-dependent mobilities}

\label{sec:lendep}

Several recent works have shown using MD simulations that the dislocation mobility in solid solutions is length-dependent if the periodic line length is too small~\cite{osetskyTwoModesScrew2019,sillsLineLengthDependentDislocationMobilities2020}.
In this length-dependent regime, the line is seen to occasionally become completely pinned during a simulation.
In contrast, a longer dislocation line moves continuously in time.
Sills et al. showed that at a given stress and temperature, a minimum line length could be identified above which the dislocation mobility was length-independent~\cite{sillsLineLengthDependentDislocationMobilities2020}.
They further showed that this minimum line length decreased with increasing stress and temperature.
A conclusive justification for these observations has yet to be presented. 
Sills et al. showed a similar length-dependence in kinetic Monte Carlo simulations based on the solid solution strengthening model of Leyson et al.~\cite{leysonQuantitativePredictionSolute2010}, but a fundamental explanation is still lacking.

Using the mobility model presented above, a physical explanation can be provided.
According to Eq.~(\ref{eq:vbar}), in order for the mobility to be length-independent, the system size must be large enough so that the mean values of $\overline{\Delta A}(l_\text{b})$ and $\bar{t}_w(l_\text{b})$ in addition to the bowing density of states $n_b(l_\text{b})$ are converged.
If the system size is too small, $\overline{\Delta A}(l_\text{b})$, $\bar{t}_w(l_\text{b})$, and $n_b(l_\text{b})$ will vary as the dislocation line glides through the simulation cell.
For example, if the dislocation happens to move to a location where $\bar{t}_w(l_\text{b})$ is rather large for all bowing lengths, then the entire dislocation line may become pinned.
In other words, the concept of a length-dependent mobility derives from a poor statistical sampling of the dislocation-solute interactions.
Future work could focus on demonstrating this effect in quantitative terms, rather than the simple qualitative argument given here.




\section*{Conclusions}

In this work we have analyzed dislocation line geometries obtained from MD data sets in an effort to better understand the fundamentals of dislocation-solute interactions and their relationship to solute strengthening and dislocation mobility in solid solution systems.
Consistent with previous studies, we find that dislocation motion through a solid solution system is well described by the theory of depinning of elastic manifolds.
In such a view, dislocation-solute interactions are categorized based on the position of the current state (stress and temperature) relative to the depinning transition, where the dislocation transitions from immobile to mobile (at $T=0$~K).
Before the depinning transition, the dislocation adopts a rough shape characterized by two different roughness profiles.
Below a correlation length $\xi$ the roughness scales as $l^{H_1}$ and above $\xi$ it scales as $l^{H_2}$.
The Hurst (roughness) exponent $H_1$ is found to vary between 0.5 and 1 ($\sigma<\sigma_\text{c}$), indicating a correlated, rough line profile.
We have interpreted this roughness as being the result of coherent line bowing over all bowing lengths below $2\xi$.
In contrast, $H_2$ varies between 0 and 0.5, corresponding to uncorrelated roughness resulting from random thermal noise and incoherent bowing induced by solute interactions.
As the stress increases so too does $\xi$, until at the depinning transition $\xi\rightarrow\infty$.
Above the depinning transition, the correlation length drops to zero and the line adopts an uncorrelated profile with $H\approx0.5$ indicating that the solute field acts as a random noise field.
These behaviors are consistent across all temperatures considered.

The basic picture of dislocation-solute interactions which emerges from this analysis is starkly different from that assumed in theories of solid solution strengthening.
Namely, we interpret the line roughness prior to depinning at scales below $\xi$ as resulting from coherent line bowing over all length scales between $l_\text{min}$ (set by the discreteness of the lattice) and $2\xi$.
This interpretation brings into question the simplifying geometric assumptions which are made in strengthening theories; primarily that the mean bowing length at zero stress $\bar{l}_\text{b0}$ governs strength. 
In an effort to develop a more rigorous formulation which is consistent with our observations, we developed a creep mobility model which invokes the notion of a bowing density of states, $n_\text{b}(l_\text{b})$.
This mobility model provides one possible route towards developing a comprehensive theory for dislocation-solute interactions in random solid solutions.
Using this model, we were able to provide a physically motivated explanation for the line-length-dependent dislocation mobilities which have been observed in several recent studies of dislocation motion in random solid solutions.

Our results demonstrate the need for more research into the fundamentals of dislocation-solute interactions which leverage modern computational tools.
Future work should seek to better quantify the fundamental quantities which govern dislocation motion before the depinning transition ($n_\text{b}(l_\text{b})$, $\overline{\Delta A}(l_\text{b})$, $\bar{t}_w(l_\text{b})$) and extend our analysis to dislocations of other character angles, solid solution systems, solute concentrations, and crystal structures (e.g., body-centered cubic, hexagonal closed-packed).
Using these advances in knowledge and building on the rich body of work preceding them, it may become possible to predict strengthening and mobility from fundamental principles of statistical physics and physical metallurgy.

\section*{Appendix A}

Here we analyze the influence of the dislocation extraction algorithm (DXA) line smoothing operation on the line geometry~\cite{stukowskiExtractingDislocationsNondislocation2010,stukowskiVisualizationAnalysisAtomistic2010}.
DXA operates by constructing a triangulation of the crystal lattice, and then using this triangulation to form Burgers circuits and localize the dislocation line.
As a result, the ``raw'' dislocation line extracted by DXA is quite jagged, since it is comprised of short line segments (on the order of a lattice constant in length) from the triangulation.
For example, even if the dislocation line itself is completely straight, the raw dislocation line identified by DXA may be jagged (depending on its orientation relative to the lattice). 
To remove this jaggedness and produce a dislocation line with a more physical geometry, DXA employs a two step process.
First, the line is ``coarsened'' by removing some of the nodes. 
And secondly, the line is ``smoothed.''
Since the analysis here is focused on the details of the dislocation line geometry, there is a concern that the coarsening and smoothing operations could introduce artifacts.

\begin{figure}[h!]
  \caption{\csentence{Comparison of roughness plots with and without coarsening and smoothing (C\&S) in DXA}
      Roughness plots with and without for C\&S at two different temperatures and stresses. \label{fig:roughness-dxa}}
  \includegraphics[width=0.8\linewidth]{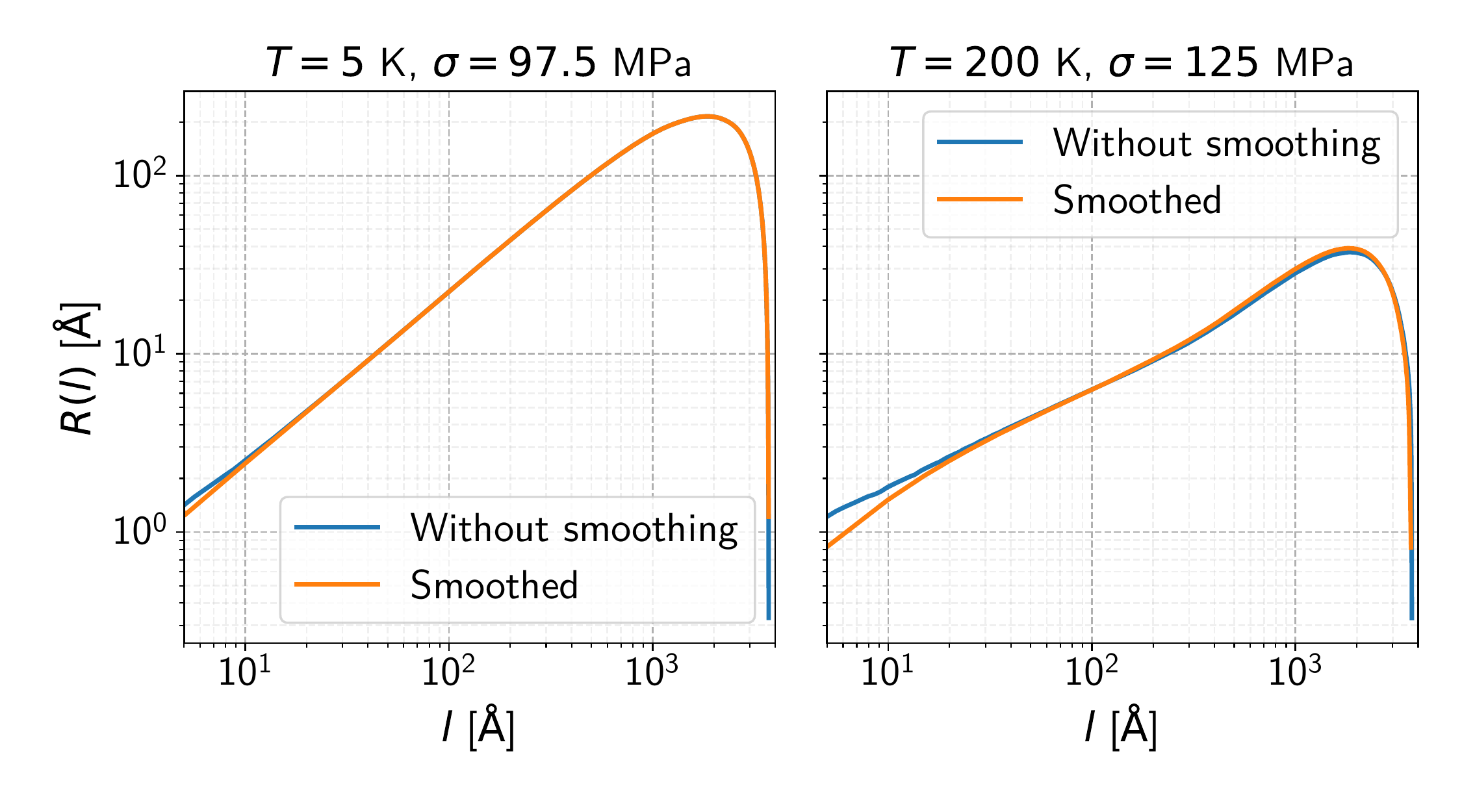}
\end{figure}

To examine the influence of coarsening and smoothing (C\&S), we show in Fig.~\ref{fig:roughness-dxa} the roughness curves at two temperatures and stresses with and without C\&S.
Fig.~\ref{fig:roughness-dxa} shows that the roughness curves are nearly identical except at small bowing lengths $l$.
Specifically, we find that the dislocation lines are rougher at bowing lengths below about 30~\AA~when C\&S is not applied.
This makes sense because the raw dislocations are quite rough at small length scales as a result of the triangulation.
One way to interpret this result is that at bowing lengths below 30~\AA, the results should be considered ``untrustwothy'' since they are apparently influenced by the details of how the dislocation lines are extracted.
A simple way to avoid this issue is then to truncate datasets below $l<30$~\AA.
When the data is truncated in this way, the main influence is that small $\xi$ values become difficult to compute reliably.
In terms of the major results presented in Fig.~\ref{fig:stress-t-d-xi}, the values of $\xi$ at high stress are affected by the data truncation.
For this reason, we indicate our uncertainty in these plots as a gray area.

It is important to note that some sort of smoothing operation is inevitable when extracting dislocations from atomistic datasets, even if the smoothing is implicit.
For example, Geslin and Rodney~\cite{geslin2018thermal} extracted dislocation lines by averaging atomic positions in the dislocation core using bins of a specified width.
In this case, the chosen bin width selects the smoothness of the dislocation lines.


\begin{backmatter}

\section*{Availability of data and materials}
All of the raw datasets on which the results of the manuscript are based are publicly available via the following Data DOI: 10.5281/zenodo.3899946.

\section*{Competing interests}
  The authors declare that they have no competing interests.
  
\section*{Funding}

This work was completed in the ELTE Institutional Excellence Program (1783-3/2018/FEKUTSRAT) supported by the Hungarian Ministry of Human Capacities. GP and PDI acknowledge the support of the National Research, Development and Innovation Fund of Hungary (contract numbers: NKFIH-K-119561, NKFIH-KH-125380).
MEF, XWZ, and RBS gratefully acknowledge the support of the U.S. Department of Energy, Office of Energy Efficiency and Renewable Energy, Fuel Cell Technologies Office and Sandia National Laboratories. 
Sandia National Laboratories is a multi-mission laboratory managed and operated by National Technology and Engineering Solutions of Sandia, LLC., a wholly owned subsidiary of Honeywell International, Inc., for the U.S. Department of Energy’s National Nuclear Security Administration under contract DE-NA-0003525.
This paper describes objective technical results and analysis. Any subjective views or opinions that might be expressed in the paper do not necessarily represent the views of the U.S. Department of Energy or the United States Government.

\section*{Author's contributions}

PDI and RBS designed and planned the study. MEF and XWZ performed and processed the MD simulations. GP, PDI, and RBS analyzed and interpreted the results. GP, PDI, and RBS prepared the manuscript.    

\section*{Acknowledgements}

Not applicable.


\bibliographystyle{bmc-mathphys} 
\bibliography{bmc_article}      



\end{backmatter}
\end{document}